\documentclass[aps,pra,twocolumn,showpacs]{revtex4}
\usepackage{bbm,amsmath,amssymb,amsbsy,graphicx,times,subfigure}

\usepackage[latin1]{inputenc}
\newcommand{\gr}[1]{\boldsymbol{#1}}
\newcommand{\be}{\begin{equation}}
\newcommand{\ee}{\end{equation}}
\newcommand{\bea}{\begin{eqnarray}}
\newcommand{\eea}{\end{eqnarray}}
\newcommand{\ket}[1]{|#1\rangle}
\newcommand{\bra}[1]{\langle#1|}
\newcommand{\avr}[1]{\langle#1\rangle}
\newcommand{\N}{{\cal N}}
\newcommand{\D}{\Delta}
\newcommand{\abs}[1]{\left\vert#1\right\vert}
\renewcommand{\det}{{\rm Det}\,}
\newcommand{\eq}[1]{Eq.~(\ref{#1})}
\newcommand{\tr}[1]{\,{\rm Tr}\,#1}

\begin{document}
\title{Extremal entanglement and mixedness in continuous variable systems}
\date{May 7, 2004}
\author{Gerardo Adesso}
\author{Alessio Serafini}
\author{Fabrizio Illuminati}
\affiliation{Dipartimento di Fisica ``E. R. Caianiello'',
Universit\`a di Salerno, INFM UdR di Salerno, INFN Sezione di Napoli,
Gruppo Collegato di Salerno,
Via S. Allende, 84081 Baronissi (SA), Italy}
\begin{abstract}
We investigate the relationship between mixedness and entanglement
for Gaussian states of continuous variable systems.
We introduce generalized entropies based on Schatten $p$-norms to quantify the mixedness of a state,
and derive their explicit expressions in terms of symplectic spectra. We
compare the hierarchies of mixedness provided by such measures with the one
provided by the purity (defined as ${\rm tr}\,\varrho^2$ for the state $\varrho$) for generic
$n$-mode states. We then review the analysis proving the existence of both maximally and minimally
entangled states at given global and marginal purities, with the entanglement quantified
by the logarithmic negativity. Based on these results, we extend such an analysis to generalized
entropies, introducing and fully characterizing maximally and minimally entangled states
for given global and local generalized entropies. We compare the different roles played by the purity
and by the generalized $p$-entropies in quantifying the entanglement and the mixedness of
continuous variable systems. We introduce the concept of average logarithmic negativity, showing
that it allows a reliable quantitative estimate of continuous variable entanglement by direct
measurements of global and marginal generalized $p$-entropies.
\end{abstract}

\maketitle

\section{Introduction}
The degree of entanglement ({\em i.e.~}the contents in {\em
quantum} correlations) as well as the degree of mixedness ({\em
i.e.~} ``the amount'' by which a state fails to be pure) are among
the crucial features of quantum states from the point of view of
quantum information theory. Indeed, the search for proper
analytical ways to quantify such features for general (mixed)
quantum states cannot be yet considered accomplished. In view of
such considerations, it is clear that the full understanding of
the relationships between the quantum correlations contained in a
bipartite state and the global and local ({\em i.e.~}referring to
the reduced states of the two subsystems) degrees of mixedness of
the state, would be desirable. In particular, it would be a
relevant step towards the clarification of the nature of quantum
correlations and, possibly, of the distinction between quantum and
classical correlations of mixed states, which remains an open
issue \cite{henvedral01}. A simple question one can raise in such
a context is the investigation of the properties of extremally
entangled states for a given degree of mixedness.\par
Let us mention that, as for two--qubit systems, the notion of
maximally entangled states at fixed mixedness (MEMS) was
originally introduced by Ishizaka and Hiroshima
\cite{hishizaka00}. The discovery of such states spurred several
theoretical works \cite{vers01}, aimed at exploring the relations
between different measures of entanglement and mixedness \cite{wei03}
(strictly related to the question of the {\em ordering} of
these different measures \cite{order}).
Moreover, maximally entangled states for given
local (or ``marginal'') mixednesses have been recently introduced and
analyzed in detail in the context of qubit systems
\cite{adesso03}. On the experimental side, much
attention has been devoted to exploring the two-qubit
Hilbert space in optical settings \cite{white02}, while the
experimental realization of MEMS has been recently demonstrated
\cite{demartonzo03}.\par
Because of the great current interest
in continuous variable (CV) quantum information \cite{pati03, crypto,
tele, dense}, the extension of such analyses to infinite
dimensional systems is higly desirable. In the present work,
we introduce and study in detail extremally entangled mixed Gaussian states
of infinite dimensional Hilbert spaces for fixed global and marginal
generalized entropies, significantly generalizing
the results derived earlier in Ref.~\cite{adeser04}, where the
existence of maximally and minimally entangled mixed Gaussian states
at given global and marginal purities was first discovered.
In the present paper we will make use of a hierarchy of generalized
entropies, based on the Schatten $p$-norms, to quantify mixedness
and characterize extremal entanglement in continuous varibale (CV)
systems and investigate several
related subjects, like the ordering of such different entropic
measures and the relations between EPR (Einstein-Podolsky-Rosen)
correlations and symplectic spectra. The crucial starting point
of our analysis is the observation that the existence of infinitely
entangled states \cite{jens02,keylwer} prevents maximally entangled
Gaussian states from being defined as states with maximal, finite,
logarithmic negativity, even at fixed global mixedness \cite{foot1}.
However, we will show that fixing the global {\em and} the local
mixednesses allows to define unambiguously both
maximally and minimally entangled mixed Gaussian states. This
relevant and somehow surprising result -- there is no analog of
Gaussian minimally entangled states in finite dimensional systems --
turns out to have an experimental
interest as well \cite{adeser04,fiurcerf03}. \par The paper is
structured as follows. In Sec.~\ref{sec2m} we briefly review
the basic notation and the general properties of
Gaussian states. In Sec.~\ref{mmix} we introduce the hierarchy of
generalized Schatten $p$-norms and entropies, and
we extensively discuss the problem of the ordering of different
entropic measures for states with an arbitrary number of modes.
In Sec.~\ref{ment} we review the state of the art on the existing,
computable measures of entanglement for two--mode Gaussian states,
while in Sec.~\ref{epr} we present a heuristic argument relating
EPR correlations to symplectic spectra. In Sec.~\ref{para}
we introduce a parametrization of two--mode Gaussian states in
terms of symplectic invariants endowed with a direct physical
interpretation. Exploiting these results, in Sec.~\ref{extre}
we define maximally and minimally entangled states for given
global and marginal purities and present some experimental
situations in which these states occur. In Sec.~\ref{genent} we generalize
the concept of extremal entanglement in CV systems by introducing
maximally and minimally entangled states for given global and local
generalized entropies of arbitrary order, and we present an extensive
study of their properties. We find out, somehow surprisingly, that
maximally and minimally entangled states can interchange their roles
for certain ranges of values of the global and marginal $p$-entropies;
moreover, we observe that with increasing $p$ the generalized entropies,
while carrying in general less information on a quantum state, provide
a more accurate quantification of its entanglement.
In Sec.~\ref{aln} we introduce and study
the concept of ``average logarithmic negativity'', showing that this
quantity provides an excellent quantitative estimate of CV entanglement
based only on the knowledge of global and local entropies.
Finally, in Sec.~\ref{concl} we summarize our results
and discuss future perspectives.
\section{Gaussian states: general overview\label{sec2m}}
Let us consider a CV system, described by an
Hilbert space ${\cal H}=\bigotimes_{i=1}^{n}
{\cal H}_{i}$ resulting from the tensor product
of the infinite dimensional Fock spaces ${\cal H}_{i}$'s.
Let $a_{i}$ be the annihilation operator acting on ${\cal H}_{i}$,
and $\hat x_{i}=(a_{i}+a^{\dag}_{i})$ and
$\hat p_{i}=(a_{i}-a^{\dag}_{i})/i$ be
the related quadrature phase operators.
The corresponding phase space variables will be denoted by $x_{i}$ and $p_{i}$.
Let us group together the operators $\hat x_{i}$ and $\hat p_{i}$ in a
vector of operators $\hat X = (\hat x_{1},\hat p_{1},\ldots,\hat x_{n},\hat p_{n})$.
The canonical
commutation relations for the $\hat X_{i}$'s are encoded in
the symplectic form $\gr{\Omega}$
\[
[\hat X_{i},\hat X_j]=2i\Omega_{ij} \; ,
\]
\[
{\rm with}\quad\boldsymbol{\Omega}\equiv \bigoplus_{i=1}^{n}
\gr{\omega}\; , \quad \boldsymbol{\omega}\equiv \left( \begin{array}{cc}
0&1\\
-1&0
\end{array}\right) \; .
\]
The set of Gaussian states is, by definition, the set of states
with Gaussian characteristic functions and quasi--probability
distributions. Therefore a Gaussian state $\varrho$ is completely
characterized by its first and second statistical moments, which
form, respectively, the vector of first moments $\bar
X\equiv\left(\langle\hat X_{1} \rangle,\langle\hat
X_{1}\rangle,\ldots,\langle\hat X_{n}\rangle, \langle\hat
X_{n}\rangle\right)$ and the covariance matrix of elements
$\boldsymbol{\sigma}$
\begin{equation}
\sigma_{ij}\equiv\frac{1}{2}\langle \hat{X}_i \hat{X}_j +
\hat{X}_j \hat{X}_i \rangle -
\langle \hat{X}_i \rangle \langle \hat{X}_j \rangle \, , \label{covariance}
\end{equation}
where, for any observable $\hat{o}$, the expectation value
$\langle\hat o\rangle\equiv\,{\rm Tr}(\varrho\hat o)$.
First statistical moments can be arbitrarily adjusted by
local unitary operations, which cannot affect any property related
to entanglement or mixedness. Therefore they will be unimportant
to our aims and we will set them to $0$ in the following, without
any loss of generality. Throughout the paper, $\gr{\sigma}$ will
stand for the covariance matrix of the Gaussian state
$\varrho$.\par Let us now consider the hermitian operator
$\hat{y}=Y\hat{X}^T$, where $Y\in\mathbb{R}^{2n}$ is an arbitrary
real $2n$-dimensional row vector. Positivity of $\varrho$ imposes
${\rm Tr}(\varrho\hat{y}^2)\ge0$, which can be simply recast in
terms of second moments as $Y\gr{\tau}Y^T\ge0 $, with
$\tau_{ij}=\langle\hat{X}_{i}\hat{X}_{j}\rangle$. From this
relation, exploiting the canonical commutation relations and
recalling definition~(\ref{covariance}) and the arbitrarity of $Y$,
the Heisenberg uncertainty principle can be recast in the form
\begin{equation}
\boldsymbol{\sigma}+ i\boldsymbol{\Omega}\ge 0 \; ,
\label{bonfide}
\end{equation}
Inequality (\ref{bonfide}) is the
necessary and sufficient constraint $\boldsymbol{\sigma}$
has to fulfill to
be a {\em bona fide} covariance matrix \cite{simon87, simon94}. We mention
that such a constraint implies $\gr{\sigma}\ge0$.\par
In the following, we will make use of the Wigner
quasi--probability representation
$W(x_{i},p_{i})$ defined, for any density matrix, as the Fourier transform of the
symmetrically ordered characteristic function \cite{barnett}.
In Wigner phase space picture, the tensor product
${\cal H}=\bigotimes{\cal H}_{i}$ of the Hilbert
spaces ${\cal H}_{i}$'s of the $n$ modes results in the direct sum
$\Gamma=\bigoplus\Gamma_{i}$ of the phase spaces
$\Gamma_{i}$'s.
In general, as a consequence of the Stone-von Neumann theorem,
any symplectic transformation acting on the
phase space $\Gamma$ corresponds to a unitary operator acting on the
Hilbert space $\cal H$, through the so called {\em metaplectic}
representation. Such a unitary operator is generated
by terms of the second order in the
field operators \cite{simon94}.
In what follows we will refer to a transformation $S_{l} = \bigoplus S_{i}$, with each
$S_{i} \in Sp_{(2,\mathbb R)}$ acting on
$\Gamma_{i}$, as to a ``local symplectic operation''.
The corresponding unitary transformation is
the ``local unitary transformation'' $U_{l}=
\bigotimes U_{i}$, with each $U_{i}$ acting on ${\cal H}_{i}$.\par
The Wigner function of a Gaussian state
can be written as follows in terms of phase space quadrature variables
\begin{equation}
W(X)=\frac{\,{\rm e}^{-\frac{1}{2}X\boldsymbol{\sigma}^{-1}X^{T}}}{\pi\sqrt{{\rm
Det}\,\boldsymbol{\sigma}}}{\:,}\label{wigner}
\end{equation}
where $X$ stands for the vector $(x_{1},p_{1},\ldots,x_{n},p_{n})\in\Gamma$. \par
Finally let us recall that, due to Williamson theorem \cite{williamson36},
the covariance matrix of a $n$--mode Gaussian state can always be
written as \cite{simon87}
\begin{equation}
\gr{\sigma}=S^T \gr{\nu} S \; , \label{willia}
\end{equation}
where $S\in Sp_{(2n,\mathbb{R})}$ and $\gr{\nu}$ is the covariance matrix
\begin{equation}
\gr{\nu}=\,{\rm diag}({\nu}_{1},{\nu}_{1},\ldots,{\nu}_{n},{\nu}_{n}) \, ,
\label{therma}
\end{equation}
corresponding to a tensor product of thermal states with diagonal
density matrix $\varrho^{_\otimes}$ given by \be
\varrho^{_\otimes}=\bigotimes_{i}
\frac{2}{\nu_{i}+1}\sum_{k=0}^{\infty}\left(
\frac{\nu_{i}-1}{\nu_{i}+1}\right)\ket{k}_{i}{}_{i}\bra{k}\; ,
\label{thermas} \ee $\ket{k}_i$ being the $k$-th number state of the
Fock space ${\cal H}_{i}$. The dual (Hilbert space) formulation
of \eq{willia} then reads: $\varrho=U^{\dag}\,\varrho^{_\otimes}\,
U$, for
some unitary $U$.\\
The quantities $\nu_{i}$'s form the symplectic spectrum of
the covariance matrix $\gr{\sigma}$ and can be
computed as the eigenvalues of the matrix $|i\gr{\Omega}\gr{\sigma}|$.
Such eigenvalues are in fact invariant under the action
of symplectic transformations on the matrix $\gr{\sigma}$.\\
The symplectic eigenvalues $\nu_{i}$ encode essential informations
on the Gaussian state $\gr{\sigma}$ and provide powerful, simple
ways to express its fundamental properties. For instance, let us
consider the Heisenberg uncertainty relation (\ref{bonfide}).
Since $S$ is symplectic, one has $S^{-1
T}\gr{\Omega}S^{-1}=\gr{\Omega}$, so that inequality
(\ref{bonfide}) is {\em equivalent} to $\gr{\nu}+i\gr{\Omega}\ge
0$. In terms of the symplectic eigenvalues $\nu_{i}$ the
uncertainty relation then simply reads \be {\nu}_{i}\ge1 \; .
\label{sympheis} \ee We can, without loss of generality, rearrange
the modes of a $n$-mode state such that the corresponding
symplectic eigenvalues are sorted in ascending order $$ \nu_1 \le
\nu_2 \le \ldots \le \nu_{n-1} \le \nu_n\,.$$ With this notation,
the uncertainty relation reduces to $\nu_1 \ge 1$. We remark that
the full saturation of Heisenberg uncertainty principle can only
be achieved by pure $n$-mode Gaussian states, for which
$\nu_i=1\,\,\forall i=1,\ldots, n$. Instead, mixed states
such that $\nu_{i\le k}=1$ and $\nu_{i>k}>1$, with $1\le k\le n$,
only partially saturate the uncertainty principle, with partial
saturation becoming weaker with decreasing $k$. Such states are
minimum uncertainty mixed Gaussian states in the sense that
the phase quadrature operators of the first $k$ modes
satisfy the Heisenberg minimal uncertainty, while for the
remaining $n-k$ modes the state indeed contains some
additional thermal and/or Schr\"odinger--like correlations
which are responsible for the global
mixedness of the state.

\subsection{Two--mode states}
In the present work we will mainly deal with two--mode
Gaussian states. Here, we will thus
briefly review their relevant properties and specify
some further notations.\par
The expression of the two--mode covariance matrix
$\boldsymbol{\sigma}$ in terms of the three $2\times 2$
matrices $\boldsymbol{\alpha}$, $\boldsymbol{\beta}$, $\boldsymbol{\gamma}$
will be useful
\begin{equation}
\boldsymbol{\sigma}\equiv\left(\begin{array}{cc}
\boldsymbol{\alpha}&\boldsymbol{\gamma}\\
\boldsymbol{\gamma}^{T}&\boldsymbol{\beta}
\end{array}\right)\, . \label{espre}
\end{equation}
It is well known that for any two--mode
covariance matrix $\boldsymbol{\sigma}$ there exists a local
symplectic operation $S_{l}=S_{1}\oplus S_{2}$ which takes
$\boldsymbol{\sigma}$ to the so called standard form $\boldsymbol{\sigma}_{sf}$
\cite{simon00, duan00}
\begin{equation}
S_{l}^{T}\boldsymbol{\sigma}S_{l}=\boldsymbol{\sigma}_{sf}
\equiv \left(\begin{array}{cccc}
a&0&c_{+}&0\\
0&a&0&c_{-}\\
c_{+}&0&b&0\\
0&c_{-}&0&b
\end{array}\right)\; . \label{stform}
\end{equation}
States whose standard form fulfills $a=b$ are said to be symmetric.
Let us recall that any pure state is symmetric and fulfills
$c_{+}=-c_{-}=\sqrt{a^2-1}$.
The correlations $a$, $b$, $c_{+}$, and $c_{-}$ are determined by the four local symplectic
invariants ${\rm Det}\boldsymbol{\sigma}=(ab-c_{+}^2)(ab-c_{-}^2)$,
${\rm Det}\boldsymbol{\alpha}=a^2$, ${\rm Det}\boldsymbol{\beta}=b^2$,
${\rm Det}\boldsymbol{\gamma}=c_{+}c_{-}$. Therefore, the standard form
corresponding to any covariance matrix is unique.  \par
The uncertainty principle Ineq.~(\ref{bonfide}) can be recast as a constraint on
the $Sp_{(4,{\mathbb R})}$ invariants ${\rm Det}\gr{\sigma}$ and
$\Delta(\gr{\sigma})={\rm Det}\boldsymbol{\alpha}+\,{\rm Det}\boldsymbol{\beta}+2
\,{\rm Det}\boldsymbol{\gamma}$:
\begin{equation}
\Delta(\gr{\sigma})\le1+\,{\rm Det}\boldsymbol{\sigma}
\label{sepcomp}\; .
\end{equation}
Let us mention that, as it is evident from \eq{stform}, the condition $\gr{\sigma}\ge 0$
implies
\be
ab-c_{\mp}^2\ge0 \; . \label{disut}
\ee

The symplectic eigenvalues of a two--mode Gaussian state will be
named $\nu_{-}$ and $\nu_{+}$, with $\nu_{-}\le \nu_{+}$ in
general. With such an ordering, the Heisenberg uncertainty
relation \eq{sympheis} becomes
\be
\nu_{\mp} \ge 1 \; .
\label{sympheisweak}
\ee

Full saturation $\nu_{-}=\nu_{+}=1$ yields the standard
pure Gaussian states of Heisenberg minimum uncertainty;
partial saturation $\nu_{-}=1$, $\nu_{+} > 1$ defines
the minimum-uncertainty mixed Gaussian states.
A simple expression for the $\nu_{\mp}$ can be found in terms of
the two $Sp_{(4,\mathbb{R})}$ invariants \cite{vidwer, serafozzi}
\begin{equation}
2{\nu}_{\mp}^2=\Delta(\gr{\sigma})\mp\sqrt{\Delta(\gr{\sigma})^2
-4\,{\rm Det}\,\gr{\sigma}} \, .
\label{sympeig}
\end{equation}

\noindent In turn, Eq.~(\ref{sympeig}) yields immediately
\begin{equation}
\Delta(\gr{\sigma}) \, = \, {\nu}_{-}^2 \, + \, {\nu}_{+}^2 \; .
\label{simplerelation}
\end{equation}

A subclass of Gaussian states that will play a relevant role in
the following is constituted by the nonsymmetric two--mode
squeezed thermal states. Let $S_{r}=\exp(\frac12 r
a_{1}a_{2}-\frac12 r a_{1}^{\dag}a_{2}^{\dag})$ be the two mode
squeezing operator with real squeezing parameter $r$, and let
$\varrho^{_\otimes}_{\nu_i}$ be a tensor product of thermal states
with covariance matrix ${\gr\nu}_{\nu_{\mp}}= {\mathbbm
1}_{2}\nu_- \oplus {\mathbbm 1}_{2}\nu_{+}$, where $\nu_{\mp}$ is,
as usual, the symplectic spectrum of the state. Then, a
nonsymmetric two-mode squeezed thermal state $\xi_{\nu_{i},r}$ is
defined as
${\xi}_{\nu_{i},r}=S_{r}\varrho^{_\otimes}_{\nu_i}S_{r}^{\dag}$,
corresponding to a standard form with \bea
a&=&{\nu_-}\cosh^{2}r+{\nu_{+}}\sinh^{2}r \; ,\nonumber\\
b&=&{\nu_{-}}\sinh^{2}r+{\nu_{+}}\cosh^{2}r \; ,\label{2mst}\\
c_{\pm}&=&\pm\frac{\nu_-+\nu_+}{2}\sinh2r \; . \nonumber
\eea
We mention that the peculiar entanglement properties of squeezed
nonsymmetric thermal states
have been recently analyzed \cite{jiang03}.
In the symmetric instance (with $\nu_-=\nu_+=\nu$)
these states reduce to two--mode squeezed thermal
states. The covariance matrices of these states are
symmetric standard forms with
\be
a=\nu\,{\cosh2r}\; , \quad
c_{\pm}=\pm\nu\,{\sinh2r} \, . \label{sqthe}
\ee
In the pure case, for which $\nu=1$, one recovers the
two--mode squeezed vacua. Notice that such states encompass
all the standard forms associated to pure states: any two--mode
Gaussian state can reduced to a squeezed vacuum by means of
unitary local operations.\\
Two--mode squeezed states (both thermal and pure)
are endowed with remarkable properties
related to entanglement \cite{2max, giedke03, rigesc03}.
Even the ideal, perfectly correlated original EPR state \cite{epr35}
can be seen indeed as a two--mode squeezed vacuum
in the limit of infinite squeezing parameter $r$.
The dynamical properties of the entanglement and the
characterization of the decoherence of such states have been
addressed in detail in several works \cite{dumodi}.
We will also show, as a byproduct of the present
work, the peculiar role they play
as maximally entangled Gaussian states.\par
\section{Measures of mixedness, degree of coherence, and entropic
measures \label{mmix}}
The degree of mixedness of a quantum state
$\varrho$ can be characterized completely by
the knowledge of all the associated Schatten $p$--norms
\cite{bathia}
\be
\|\varrho\|_p\equiv(\,{\rm
Tr}\,|\varrho|^p)^{\frac1p} =(\,{\rm Tr}\,\varrho^p)^{\frac1p}\, ,
\quad \,{\rm with} \: p \ge 1.
\ee
In particular, the case $p=2$ is
directly related to the purity $\mu=\,{\rm
Tr}\,\varrho^2= (\|\varrho\|_{2})^{2}$ \cite{paris}. The $p$-norms
are multiplicative on tensor product states and thus determine a
family of ``generalized entropies'' $S_{p}$ \cite{bastiaans,tsallis}, defined
as
\be
S_{p} = \frac{1-\,{\rm Tr}\,\varrho^p}{p-1} \; , \quad p > 1.
\label{pgen}
\ee
These quantities have been introduced independently by M. J. Bastiaans
in the context of quantum optics \cite{bastiaans}, and by C. Tsallis
in the context of statistical mechanics \cite{tsallis}. In the
quantum arena, they can be interpreted both as quantifiers of
the degree of mixedness of a state $\varrho$ by
the amount of information it lacks, and as measures of the overall degree
of coherence of the state (the latter meaning was elucidated by
Bastiaans in his analysis of the properties of partially coherent
light). The quantity
$S_{2}=1-\mu\equiv S_L$, conjugate to the purity $\mu$, is usually
referred to as the linear entropy: it is a particularly important measure
of mixedness, essentially because of the simplicity of its analytical
expressions, which will become soon manifest. Finally, another
important class of entropic measures includes the R\'{e}nyi
entropies \cite{renyi}
\be
S_{p}^{R} = \frac{\ln \, {\rm Tr} \, \varrho^{p}}{1-p} \; , \quad p > 1.
\ee
It can be easily shown that
\be
\lim_{p\rightarrow1+}S_{p}=\lim_{p\rightarrow1+}S_{p}^{R}=
-\,{\rm Tr}\,(\varrho\ln\varrho) \equiv S_{V} \, ,
\label{genvneu}
\ee
so that also the Shannon-von Neumann
entropy $S_{V}$ can be defined in terms of $p$-norms. The quantity
$S_{V}$ is additive on tensor product states and provides a
further convenient measure of mixedness of the quantum state
$\varrho$.\par It is easily seen that the generalized entropies
$S_p$'s range from $0$ for pure states to $1/(p-1)$ for
completely mixed states with fully degenerate eigenspectra.
Notice that $S_V$ is infinite on infinitely mixed states, while
$S_L$ is normalized to $1$. We also mention that, in the
asymptotic limit of arbitrary large $p$, the function
$\tr{\varrho^p}$ becomes a function only of the largest eigenvalue
of $\varrho$: more and more information about the state is
discarded in such an estimate for the degree of purity;
considering $S_p$ in the limit $p\rightarrow \infty$
yields a trivial constant null function, with no information at
all about the state under exam. We also note that, for any given
quantum state, $S_{p}$ is a monotonically decreasing function of $p$. \par
Because of their unitarily invariant nature, the generalized
purities ${\rm Tr}\,\varrho^p$ of generic $n$--mode Gaussian
states can be simply computed in terms of the symplectic
eigenvalues $\nu_{i}$ of $\gr{\sigma}$. In fact, a symplectic
transformation acting on $\gr{\sigma}$ is embodied by a unitary
(trace preserving) operator acting on $\varrho$, so that ${\rm
Tr}\,\varrho^{p}$ can be easily computed on the diagonal state
$\nu$. One obtains \be {\rm
Tr}\,\varrho^{p}=\prod_{i=1}^{n}g_{p}(\nu_i)\; , \label{pgau} \ee
where
\[
g_{p}(x)=\frac{2^p}{(x+1)^p-(x-1)^p} \, .
\]
A first consequence of \eq{pgau} is that \be
\mu(\varrho)=\frac{1}{\prod \nu_{i}}=\frac{1}{\sqrt{{\rm
Det}\,\gr{\sigma}}} \, . \label{purgau} \ee Regardless of the
number of modes, the purity of a Gaussian state is fully
determined by the symplectic invariant ${\rm Det}\,\gr{\sigma}$
alone. A second consequence of \eq{pgau} is that, together with
Eqs.~(\ref{pgen}) and (\ref{genvneu}), it allows for the
computation of the von Neumann entropy $S_{V}$ of a Gaussian state
$\varrho$, yielding \be S_{V}(\varrho)=\sum_{i=1}^{n}f(\nu_{i}) \;
, \label{vneugau} \ee where
\[
f(x) \equiv \frac{x+1}{2}\ln\left(\frac{x+1}{2}\right)-
\frac{x-1}{2}\ln\left(\frac{x-1}{2}\right) \, .
\]
Such an expression for the von Neumann entropy of a Gaussian state
was first explicitly given in Ref.~\cite{holevo99}. Let us remark
that, clearly, the symplectic spectrum of single mode Gaussian
states, which consists of only one eigenvalue $\nu_1$, is fully
determined by the invariant ${\rm Det}\,\gr{\sigma}=\nu_{1}^2$
Therefore, all the entropies $S_{p}$'s (and $S_{V}$ as well) are
just increasing functions of ${\rm Det}\,\gr{\sigma}$
(\emph{i.e.}~of $S_L$) and induce \emph{the same} hierarchy of
mixedness on the set of one--mode Gaussian states. This is no
longer true for multi--mode states, even for the relevant, simple
instance of two--mode states.\par Here we aim to find extremal values of $S_{p}$ (for $p\neq 2$)
for fixed $S_{L}$
in the general $n$--mode Gaussian instance,
in order to quantitatively compare the characterization of mixedness
given by the different entropic measures. For simplicity,
in calculations we will replace $S_{L}$ with $\mu$.
In view of Eqs.~(\ref{pgau}) and (\ref{purgau}),
the possible values taken by $S_{p}$ for a given $\mu$
are determined by
\bea
(p-1)S_{p}=1-\left(\prod_{i=1}^{n-1}g_{p}(s_i)\right)
g_{p}\left(\frac{1}{\mu \prod_{i=1}^{n-1}s_i}\right) \, ,\label{iter}\\
{\rm with}\quad 1 \le s_i \le\frac{1} {\mu\prod_{i\neq j}s_{j}}\,
. \label{domi} \eea The last constraint on the $n-1$ real
auxiliary parameters $s_i$ is a consequence of the uncertainty
relation (\ref{sympheis}). We first focus on the instance $p<2$,
in which the function $S_{p}$ is concave with respect to any
$s_{i}$, for any value of the $s_i$'s. Therefore its minimum with
respect to, say, $s_{n-1}$ occurs at the boundaries of the domain,
for $s_{n-1}$ saturating inequality (\ref{domi}). Since $S_{p}$
takes the same value at the two extrema and exploiting
$g_{p}(1)=1$, one has \be (p-1)\min_{s_{n-1}}S_{p}=1-
\left(\prod_{i=1}^{n-2}g_{p}(s_i)\right) g_{p}\left(\frac{1}{\mu
\prod_{i=1}^{n-2}s_i}\right) \: . \ee Iterating this procedure for
all the $s_i$'s leads eventually to the minimum value
$S_{p\,min}(\mu)$ of $S_p$ at given purity $\mu$, which simply
reads \be\label{spmin}
S_{p\,min}(\mu)=\frac{1-g_p\left(\frac{1}{\mu}\right)}{p-1}\,
,\quad p<2 \, . \ee For $p<2$, the mixedness of the states with
minimal generalized entropies at given purity is therefore
concentrated in one quadrature: the symplectic spectrum of such
states is partially degenerate, with $\nu_{1}=\ldots=\nu_{n-1}=1$
and $\nu_{n}=1/\mu$. \par The maximum value $S_{p\,max}(\mu)$ is
achieved by states satisfying the coupled trascendental equations
\be g_p\left(\frac{1}{\mu\prod s_i}\right) g'_{p}(s_j)=
\frac{1}{\mu s_{j}\prod s_i}\, g_p(s_j) g'_p\left(\frac{1}{\mu
\prod s_i}\right)\, , \ee where all the products $\prod$ run over
the index $i$ from $1$ to $n-1$, and \be
 g'_p(x)=\frac{-p\,2^p\left[(x+1)^{p-1}-(x-1)^{p-1}\right]}
 {\left[(x+1)^p-(x-1)^p\right]^2}\, .
\ee It is promptly verified that the above two conditions are
fulfilled by states with a completely degenerate symplectic
spectrum: $\nu_{1}=\ldots=\nu_{n}={\mu}^{-1/n}$, yielding
\be\label{spmax} S_{p\,max}(\mu)=\frac{1-
g_p\left(\mu^{-\frac{1}{n}}\right)^n}{p-1}\, ,\quad p<2 \, . \ee

\begin{figure}[tb!]
\begin{center}
\includegraphics[width=7.5cm]{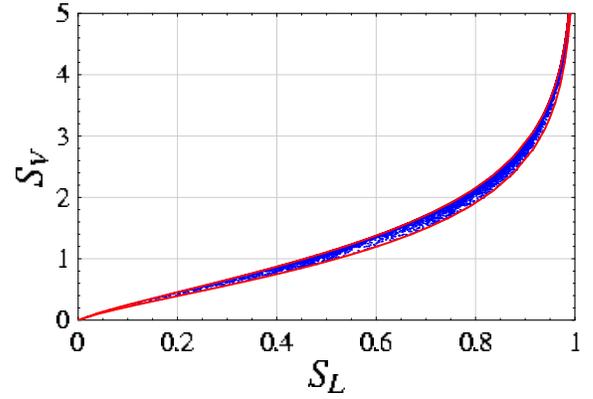}
\caption{Plot of the curves (solid lines) of maximal and minimal
von Neumann entropy at given linear entropy for two--mode Gaussian
states. The density of states in the plane of entropies is
represented as well, by plotting the distribution of 20000
randomly generated states (dots).}
\label{svvssl}
\end{center}
\end{figure}

The analysis that we carried out for $p<2$
can be straightforwardly extended to the
limit $p\rightarrow 1$, yielding the extremal values of the von Neumann
entropy for given purity $\mu$ of $n$--mode Gaussian states. Also in
this case
the states with maximal $S_{V}$ are those with a
completely degenerate symplectic spectrum, while the states with minimal $S_{V}$
are those with all the mixedness concentrated in one quadrature. The extremal values
$S_{V min}(\mu)$ and $S_{V max}(\mu)$ read
\bea
S_{V min}(\mu)&=&f\left(\frac{1}{\mu}\right) \; , \\
&&\nonumber\\
S_{V max}(\mu)&=&nf\left(\mu^{-\frac1n}\right) \; .
\eea
The behaviors of the von Neumann and of the linear entropies
for two--mode states are compared in Fig.~\ref{svvssl}.\par

The instance $p>2$ can be treated in the same way, with the major
difference that the function $S_{p}$ of \eq{iter} is convex with
respect to any $s_i$ for any value of the $s_i$'s. As a
consequence we have an inversion of the previous expressions: for
$p>2$, the states with minimal $S_{p\, min}(\mu)$ at given purity
$\mu$ are those with a fully distributed symplectic spectrum, with
\be S_{p\,min}(\mu)=\frac{1-
g_p\left(\mu^{-\frac{1}{n}}\right)^n}{p-1}\, ,\quad p>2 \, . \ee
On the other hand, the states with maximal $S_{p\,max}$ at given
purity $\mu$ are those with a spectrum of the kind
$\nu_{1}=\ldots=\nu_{n-1}=1$ and $\nu_{n}=1/\mu$. Therefore \be
S_{p\,max}(\mu)=\frac{1-g_p\left(\frac{1}{\mu}\right)}{p-1}\,
,\quad p>2 \, . \label{spglems} \ee
\par
\begin{figure}[tb!]
\begin{center}
\includegraphics[width=7cm]{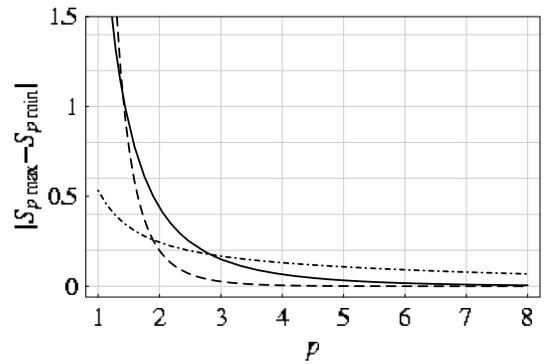}
\caption{Plot of the absolute difference between the maximal and the minimal
values of the generalized entropies $S_p$ at fixed linear entropy
$S_L$ for two--mode Gaussian states, as a function of $p$. Different curves
correspond to different values of the linear entropy:
$S_L=0.8$ (dashed line), $S_L=0.5$ (continuous line), and $S_L=0.1$
(dash-dotted line).}
\label{DeltaSvsP}
\end{center}
\end{figure}

As shown in Fig.~\ref{DeltaSvsP}, the distance
$|S_{p\,max}-S_{p\,min}|$ decreases with increasing $p$.
This is due to the fact that the quantity $S_{p}$ carries less
information with increasing $p$, and the knowledge of $\mu$
provides a more precise bound on the value of $S_{p}$.\par
\section{Measures of Entanglement\label{ment}}
We now review the main aspects of entanglement for CV systems
and discuss some of the possible quantifications
of quantum correlations for Gaussian states.\par The necessary and
sufficient separability criterion for a two--mode Gaussian state
$\varrho$ has been shown to be positivity of the partially
transposed state $\tilde{\varrho}$ (PPT criterion)
\cite{simon00}. In general, the partial transposition
$\tilde{\varrho}$ of a bipartite quantum state $\varrho$ is
defined as the result of the transposition performed on only one
of the two subsystems in some given basis. Even though the
resulting $\tilde{\varrho}$ does depend on the choice of the
transposed subsystem and on the transposition basis, the statement
$\tilde{\varrho}\ge0$ is geometric, and is invariant under such
choices \cite{perhor}. It can be easily seen from the definition
of $W(X)$ that the action of partial transposition amounts, in
phase space, to a mirror reflection of one of the four canonical
variables. In terms of $Sp_{(2,\mathbb{R})}\oplus
Sp_{(2,\mathbb{R})}$ invariants, this reduces to a sign flip in
${\rm Det}\,\gr{\gamma}$
\[
{\rm Det}\,\gr{\gamma}\;\;
\xrightarrow{\,\,\varrho\rightarrow\tilde{\varrho}\,\,} \;\;-{\rm
Det}\,\gr{\gamma}\; .
\]
Therefore
the invariant $\Delta(\gr{\sigma})$ is changed into $\tilde{\Delta}({\gr\sigma})
=\Delta(\tilde{\gr{\sigma}})=\,{\rm Det}\,\gr{\alpha}+
\,{\rm Det}\,\gr{\beta}-2\,{\rm Det}\,\gr{\gamma}$. Now, the symplectic
eigenvalues $\tilde{\nu}_{\mp}$ of $\tilde{\gr{\sigma}}$ read
\be
\tilde{\nu}_{\mp}=
\sqrt{\frac{\tilde{\Delta}(\gr{\sigma})\mp\sqrt{\tilde{\Delta}(\gr{\sigma})^2
-4\,{\rm Det}\,\gr{\sigma}}}{2}} \, . \label{sympareig}
\ee
The PPT criterion thus reduces to a simple inequality that must
be satisfied by the smallest symplectic eigenvalue $\tilde{\nu}_{-}$
of the partially transposed state
\be
\tilde{\nu}_{-}\ge 1 \: ,
\label{symppt}
\ee
which is equivalent to
\be
\tilde{\Delta}(\gr{\sigma})\le \,{\rm Det}\,\gr{\sigma}+1 \; .
\label{ppt}
\ee
The above inequalities
imply ${\rm Det}\,\gr{\gamma}=c_{+}c_{-}<0$
as a necessary condition for a two--mode Gaussian state
to be entangled.
The quantity $\tilde{\nu}_{-}$ encodes all the qualitative characterization of
the entanglement for arbitrary (pure or mixed) two--modes Gaussian states.
Note that
$\tilde{\nu}_{-}$ takes a particularly simple form for
entangled symmetric states, whose standard
form has $a=b$
\be
\tilde{\nu}_{-}=\sqrt{(a-|c_{+}|)(a-|c_{-}|)} \; .
\label{symeig}
\ee
\par
Inserting Eqs.~(\ref{2mst}) into \eq{ppt} yields the following
condition for a two-mode squeezed thermal state $\xi_{\nu_{i},r}$
to be entangled \be \sinh^2(2r) >
\frac{(\nu_+^2-1)(\nu_-^2-1)}{(\nu_-+\nu_+)^2} \; .
\label{2mstppt} \ee

As for the quantification of entanglement, no fully satisfactory
measure is known at present for arbitrary mixed bipartite Gaussian
states. However, a quantification of entanglement which can be
computed for general Gaussian states is provided by the negativity
$\N$, first introduced in Ref.~\cite{zircone}, later thoroughly discussed
and extended in Ref.~\cite{vidwer} to CV systems. The negativity
of a quantum state $\varrho$ is defined as \be {\cal
N}(\varrho)=\frac{\|\tilde \varrho \|_1-1}{2}\: , \ee where
$\tilde\varrho$ is the partially transposed density matrix and
$\|\hat o\|_1=\,{\rm Tr}|\hat o|$ stands for the Schatten $1$-norm
(the so called `trace norm') of the hermitian operator $\hat o$.
The quantity ${\cal N} (\varrho)$ is equal to
$|\sum_{i}\lambda_{i}|$, the modulus of the sum of the negative
eigenvalues of $\tilde\varrho$, quantifying the extent to which
$\tilde\varrho$ fails to be positive. Strictly related to $\N$ is
the logarithmic negativity $E_{\N}$, defined as $E_{\N}\equiv
\ln\|\tilde{\varrho}\|_{1}$, which constitutes an upper bound to
the {\em distillable entanglement} of the quantum state
$\varrho$.\par The negativity has been proven to be convex and
monotone under LOCC (local operations and classical
communications) \cite{negnote}, but fails to be continuous in
trace norm on infinite dimensional Hilbert spaces. However, this
problem can be circumvented by restricting to physical states
with finite mean energy \cite{jens02}.\par
We will now show that for any
two--mode Gaussian state $\varrho$ the negativity is a simple
decreasing function of $\tilde{\nu}_{-}$, which is thus itself an
entanglement monotone. The norm $\|\tilde\cdot\|_1$
is unitarily invariant; in particular, it is invariant under
global symplectic operations in phase space. Considering the
symplectic diagonalization $\gr{\tilde\nu}\equiv\,{\rm
diag}\,(\tilde\nu_-,\tilde\nu_-,\tilde\nu_+,\tilde\nu_+)$ of
$\tilde{\gr{\sigma}}$, this means that
$\N(\varrho)=(\|\tilde\nu\|_{1}-1)/2$. Now, because of the
multiplicativity of the norm $\|\cdot\|_1$, we have just to
compute the trace of the single--mode thermal--like operator
$\rho^{_\otimes}_{\tilde{\nu}_i}$
$$
\rho^{_\otimes}_{\tilde{\nu}_i}=\frac{2}{\tilde{\nu}_i+1}\sum_{k=0}^{\infty}
\left(\frac{\tilde{\nu}_i-1}{\tilde{\nu}_i+1}\right)^{k}\ket{k}\bra{k} \; .
$$
Such an operator is obviously normalized for $\tilde{\nu}_i\ge1$,
yielding $\|\rho^{_\otimes}_{\tilde{\nu}_i}\|_{1}=1$ (all
eigenvalues are positive). Instead, if $\tilde{\nu}_i<1$, then
$\|\rho^{_\otimes}_{\tilde{\nu}_i}\|_{1}=1/\tilde{\nu}_i$. The
proof is completed by showing that the largest symplectic
eigenvalue $\tilde \nu_{+}$ of $\tilde{\gr{\sigma}}$ fulfills
$\tilde \nu_{+}\ge 1$. This is obviously true for separable
states, therefore we can set $c_{+}c_{-}<0$. With this
position, it is easy to show that
$$
\tilde{\nu}_{+}>{\nu}_{-}\ge1 \; .
$$
Thus, we obtain $\|\rho^{_\otimes}_{\tilde{\nu}_+}\|_{1}=1$ and
\be
\|\tilde\varrho\|_{1}=\frac{1}{\tilde{\nu}_{-}}\;\Rightarrow
\N(\varrho)=\max \, \left[ 0,\frac{1-\tilde{\nu}_{-}}{2\tilde{\nu}_{-}}
\right]
\, ,
\ee
\be E_{\N}(\varrho)=\max\,\left[0,-\ln
\tilde{\nu}_{-}\right] \, .
\ee
This is a decreasing function of
the smallest partially transposed symplectic eigenvalue
$\tilde{\nu}_{-}$, quantifying the amount by which inequality
(\ref{symppt}) is violated. The eigenvalue
$\tilde{\nu}_{-}$ thus completely qualifies and quantifies the quantum
entanglement of a Gaussian state $\gr{\sigma}$. Notice that, in such
an instance, this feature holds for nonsymmetric states as well.\par
We finally mention that, as far as symmetric states
are concerned, another measure of entanglement, the entanglement
of formation $E_{F}$ \cite{bennet96}, can be actually computed
\cite{giedke03}. Since $E_{F}$ turns out to be, again, a
decreasing function of $\tilde{\nu}_{-}$, it provides for
symmetric states a quantification of entanglement fully equivalent
to the one provided by the logarithmic negativity $E_{\N}$.
\section{Symplectic eigenvalues and EPR correlations\label{epr}}
A deeper insight on the relationship between correlations and
the eigenvalue $\tilde{\nu}_{-}$
is provided by the following observation,
which holds in the symmetric case.\par
Let us define the EPR correlation $\xi$ \cite{giedke03, rigesc03}
of a continuous variable two--mode quantum state as
\be
\xi\equiv\frac{\delta_{\hat{x}_{1}-\hat{x}_{2}}+
\delta_{\hat{p}_{1}+\hat{p}_{2}}}{2}
=
\frac{{\rm Tr}\,\gr{\sigma}}{2}-\sigma_{13}
+\sigma_{24} \, ,
\ee
where $\delta_{\hat{o}}=\avr{\hat o^2}-\avr{\hat o}^2
$ for an operator $\hat o$. If $\xi\ge 1$ then the state does
not possess nonlocal correlations .
The idealized EPR-like state \cite{epr35} (simultaneous eigenstate of the commuting
observables $\hat x_{1}-\hat x_{2}$ and $\hat p_{1}+\hat p_{2}$)
has $\xi=0$.
As for standard form states, one has
\bea
\delta_{\hat{x}_{1}-\hat{x}_{2}}&=&a+b-2c_{+}\; ,\\
\delta_{\hat{p}_{1}+\hat{p}_{2}}&=&a+b+2c_{-}\; ,\\
\xi&=&a+b-c_{+}+c_{-}\; .
\eea
Notice that $\xi$ is not by itself
a good measure of correlation becasue, as one can easily verify, it is
not invariant under local symplectic operations. In particular,
applying local squeezings with parameters $r_{i}=\ln v_{i}$ and
local rotations with angles $\varphi_{i}$ to a standard form state,
we obtain \be \xi_{v_{i},\vartheta}=
\frac{a}{2}(v_{1}^{2}+\frac{1}{v_{1}^{2}})+
\frac{b}{2}(v_{2}^{2}+\frac{1}{v_{2}^{2}})
-(c_{+}v_{1}v_{2}-\frac{c_{-}}{v_{1}v_{2}})\cos{\vartheta} \, ,
\ee with $\vartheta=\varphi_{1}+\varphi_{2}$. Now, the quantity
\[
\bar\xi\equiv\min_{v_{i},\vartheta}\xi_{v_{i},\vartheta}
\]
has to be $Sp_{(2,\mathbb{R})}\oplus Sp_{(2,\mathbb{R})}$ invariant. It
corresponds to the maximal amount of EPR correlations which can be
distributed in a two-mode Gaussian state by means of local
operations. Minimization in terms of $\vartheta$ is immediate,
yielding $\bar \xi=\min_{v_{i}}\xi_{v_{i}}$, with
\be
\xi_{v_{i}}=
\frac{a}{2}(v_{1}^{2}+\frac{1}{v_{1}^{2}})+
\frac{b}{2}(v_{2}^{2}+\frac{1}{v_{2}^{2}})
-|c_{+}v_{1}v_{2}-\frac{c_{-}}{v_{1}v_{2}}| \, .
\ee
The gradient of
such a quantity is null if and only if
\bea
a \left(v_{1}^{2}-\frac{1}{v_{1}^{2}}\right)-|c_{+}|v_{1}v_{2}-
\frac{|c_{-}|}{v_{1}v_{2}}&=&0 \, ,
\label{grad1} \\
b \left(v_{2}^{2}-\frac{1}{v_{2}^{2}}\right)-|c_{+}|v_{1}v_{2}-
\frac{|c_{-}|}{v_{1}v_{2}}&=&0 \, ,
\label{grad2}
\eea
where we introduced the position $c_{+}c_{-}<0$, necessary to have
entanglement. Eqs.~(\ref{grad1}, \ref{grad2}) can be combined to
get \be a\left(v_{1}^{2}-\frac{1}{v_{1}^{2}}\right)=
b\left(v_{2}^{2}-\frac{1}{v_{2}^{2}}\right) \; . \label{conda} \ee
Restricting to the symmetric ($a=b$) entangled ($\Rightarrow
c_{+}c_{-}<0$) case, Eq.~(\ref{conda}) and the fact that $v_{i}>0$
imply $v_{1}=v_{2}$. Under such a constraint, minimizing $\xi_{v_{i}}$
becomes a trivial matter and yields
\be
\bar\xi=2
\sqrt{(a-|c_{+}|)(a-|c_{-}|)}=2 \tilde{\nu}_{-} \; .
\ee
We thus see that the smallest symplectic eigenvalue of the partially
transposed state
is endowed with a direct physical interpretation: it quantifies
the greatest amount of EPR correlations which can be created in a
Gaussian state by means of local operations. \par As can be easily
shown by a numerical analysis, such a simple interpretation is
lost for nonsymmetric states. This fact properly exemplifies the
difficulties of handling optimization problems in nonsymmetric
instances, encountered, {\em e.g.}~in the computation of the
entanglement of formation of such states.

\section{Parametrization of Gaussian states with symplectic invariants\label{para}}

Two--mode Gaussian states can be classified according to the
values of their four $Sp_{(2,\mathbb{R})}\oplus
Sp_{(2,\mathbb{R})}$ invariants $a$, $b$, $c_+$ and $c_-$, which
determine their standard form. It is relevant to provide a
reparametrization of standard form states in terms of invariants
which admit a direct interpretation for generic Gaussian states.
Such invariants will be the global purity $\mu=\tr{\varrho^2}$,
the marginal purities of the reduced states $\mu_{i}=\,{\rm
Tr}_i[(\,{\rm Tr}_{j\neq i} \varrho)^2]$ and the
$Sp_{(4,\mathbb{R})}$ invariant $\Delta$, whose meaning will
become soon clear. For a two-mode Gaussian state one has
\begin{eqnarray}
  \mu_1 &=& \frac{1}{a}\,, \quad \mu_2 \,\,=\,\, \frac{1}{b}\,, \label{mu12}\\
  \frac{1}{\mu^2} &=& \det{\gr{\sigma}} = (a b)^2-a b (c_+^2+c_-^2)+(c_+ c_-)^2 \, ,
  \label{dets}\\
  \Delta &=& a^2+b^2+2c_+ c_-\,. \label{deltas}
\end{eqnarray}
Eqs.~(\ref{mu12}-\ref{deltas}) can be inverted to provide the
following parametrization
\begin{eqnarray}
  a &=& \frac{1}{\mu_1}\,, \quad b \,\,=\,\, \frac{1}{\mu_2}\,, \label{gab} \\
c_{\pm}&=&\frac14 \sqrt{\mu_1 \mu_2 \left[\left(\D-\frac{(\mu_1 - \mu_2)^2}
{\mu_1^2 \mu_2^2}\right)^2-\frac{4}{\mu^2}\right]} \; \pm \, \epsilon \nonumber \\
\label{gc}
\end{eqnarray}
$$
{\rm with}\quad
\epsilon \, \equiv \,
\frac14 \sqrt{\frac{\left[(\mu_1+\mu_2)^2
- \mu_1^2 \mu_2^2 \D\right]^2}{\mu_1^3 \mu_2^3}
-\frac{4\mu_1 \mu_2}{\mu^2}}\, .
$$
The global and marginal purities range from $0$ to $1$,
constrained by the condition
\begin{equation}
\label{ineqmumui}
\mu \ge \mu_1 \mu_2 \; ,
\end{equation}
that can be easily shown to be a direct consequence of inequality
(\ref{disut}).
Notice that inequality (\ref{ineqmumui})
entails that no Gaussian LPTP (less pure than product) states
exist, at variance with the case of two--qubit systems
\cite{adesso03}.\par
The smallest symplectic
eigenvalues of the covariance matrix $\gr{\sigma}$ and of its
partial transpose $\tilde{\gr{\sigma}}$ are promptly determined
in terms of symplectic invariants
\begin{equation}
  2\nu_{-}^2 = \Delta-\sqrt{\Delta^2
-{\frac{4}{\mu^2}}}\,, \quad
  2\tilde{\nu}_{-}^2 = \tilde{\Delta}-\sqrt{\tilde{\Delta}^2
-{\frac{4}{\mu^2}}}\,, \label{n1}
\end{equation}
where $\tilde{\Delta} = -\D + 2/\mu_1^2 + 2/\mu_2^2$.\par The
parametrization provided by Eqs.~(\ref{gab}, \ref{gc}) describes
physical states if the radicals in Eqs.~(\ref{gc}, \ref{n1}) exist
and the Heisenberg uncertainty principle (\ref{sympheis}) is
satisfied. All these conditions can be combined and recast as
upper and lower bounds on the global symplectic invariant $\D$
\begin{eqnarray}
  & & \frac{2}{\mu} + \frac{(\mu_1 - \mu_2)^2}{\mu_1^2 \mu_2^2}
  \,\, \le\,\, \D \nonumber\\
  &\le& \min \left\{ \frac{(\mu_1 + \mu_2)^2}{\mu_1^2 \mu_2^2}
  - \frac{2}{\mu} \; , \; 1+\frac{1}{\mu^2}  \right\}
  \, . \label{deltabnd}
\end{eqnarray}
Let us investigate the role played by the invariant
$\D$ in the characterization of the properties of
Gaussian states. To this aim, we just analyse the dependence
of the eigenvalue $\tilde{\nu}_-$ on $\D$
\begin{equation}
  \left. \frac{\partial\ \tilde{\nu}^2_{-}}{\partial\ \D}
  \right|_{\mu_1,\,\mu_2,\,\mu} \,
 = \; \frac12 \left(
\frac{\tilde{\Delta}}{\sqrt{\tilde{\Delta}^2
-{\frac{1}{4 \mu^2}}}} -1 \right) \,
> 0 \, .
\label{derivata}\end{equation} The smallest symplectic eigenvalue
of the partially transposed state is strictly monotone in $\D$.
Therefore the entanglement of a generic Gaussian state
$\gr{\sigma}$ with global purity $\mu$ and marginal purities
$\mu_{1,2}$ strictly increases with decreasing $\D$. The invariant
$\D$ is thus endowed with a direct physical interpretation: at
given global and marginal purities, it determines the amount of
entanglement of the state.\par
Because, due to inequality (\ref{deltabnd}), $\D$ possess both lower and
upper bounds, not only maximally but also
\emph{minimally} entangled Gaussian states exist. This elucidates
the relations between the entanglement and the purity of two--mode
Gaussian states: the entanglement of such states is tightly bound
by the amount of global and marginal purities, with only one
remaining degree of freedom related to the invariant $\D$.

\section{Extremal entanglement at fixed global and local purities \label{extre}}

We now aim to characterize extremal (maximally and minimally)
entangled Gaussian states for fixed global and marginal purities.
As it is clear from Eq.~(\ref{mu12}), the standard form of states
with fixed marginal purities always satisfies $a=1/\mu_1$,
$b=1/\mu_2$. Therefore the complete characterization of maximally
and minimally entangled states is achieved by specifying the expression of their
standard form coefficients $c_{\mp}$.\par
Let us first consider the states
saturating the lower bound in Eq.~(\ref{deltabnd}),
which entails \emph{maximal} entanglement.
They are Gaussian maximally entangled states for fixed global and local
purities (GMEMS), admitting the
following parametrization
\begin{equation} \label{gnsm}
c_{\pm}= \pm \sqrt{\frac{1}{\mu_1 \mu_2}-\frac{1}{\mu}} \; .
\end{equation}
It is easily seen that such states belong to the class of
asymmetric two--mode squeezed thermal states, with squeezing
parameter and symplectic spectrum given by \be\label{2sqp}
\tanh2r=2(\mu_1\mu_2-\mu_1^2\mu_2^2/\mu)^{1/2}/(\mu_1+\mu_2)\, ,
\ee \be
\nu_{\mp}^2=\frac{1}{\mu}+\frac{(\mu_1-\mu_2)^2}{2\mu_1^2\mu_2^2}\mp
\frac{|\mu_1-\mu_2|}{2\mu_1\mu_2}\sqrt{\frac{(\mu_1-\mu_2)^2}{\mu_1^2\mu_2^2}
+\frac{4}{\mu}}\, . \ee In particular, any GMEMS can be written as
an entangled two-mode squeezed thermal states [satisfying
Ineq.~(\ref{2mstppt})]. This provides a characterization of
two-mode thermal squeezed states as maximally entangled states for
given global and marginal purities. We can restate this result as
follows: given an initial tensor product of (generally different)
thermal states, the unitary operation providing the maximal
entanglement for given values of the local purities $\mu_i$'s is
given by a two-mode squeezing, with squeezing parameter determined
by \eq{2sqp}. Notice that fixing the values of the local purities
is necessary to attain maximal, finite entanglment; in fact,
fixing only the value of the global purity $\mu$ always allows for
an arbitrary large logarithmic negativity (as $\tilde{\nu}_-$ goes
to $0$), achievable by means of two--mode squeezing with arbitrary
large squeezing parameter.\par Nonsymmetric two--mode thermal
squeezed states turn out to be \emph{separable} in the range
\begin{equation}
\label{gnsmsep}
\mu \le \frac{\mu_1 \mu_2}{\mu_1 + \mu_2 - \mu_1 \mu_2} \, .
\end{equation}
In such a {\it separable region} in the space of purities, no
entanglement can occur for states of the form of Eq.~(\ref{gnsm}),
while, outside this region, they are properly GMEMS. We remark
that, as a consequence, all Gaussian states whose purities fall in
the separable region defined by inequality (\ref{gnsmsep}) are
separable.\par
We now consider the states that saturate the upper
bound in Eq.~(\ref{deltabnd}). They determine the class of
Gaussian least entangled states for given global and local
purities (GLEMS). Violation of inequality (\ref{gnsmsep}) implies
that
$$
1 + \frac{1}{\mu^2} \, \le \,
\frac{(\mu_1 + \mu_2)^{2}}{\mu_1^{2} \mu_2^{2}} - \frac{2}{\mu} \; .
$$
Therefore, outside the separable region, GLEMS fulfill
\begin{equation}
\label{glm}
\D = 1+\frac{1}{\mu^2} \; .
\end{equation}
Eq.~(\ref{glm}) expresses partial saturation of Heisenberg relation
\eq{sympheisweak}. Namely, considering the symplectic diagonalization
of Gaussian states and Eq.~(\ref{simplerelation}), it immediately follows
that the $Sp_{(4,\mathbb{R})}$ invariant condition~(\ref{glm}) is fulfilled if
and only if the symplectic spectrum of the state takes the form
$\nu_-=1$, $\nu_+=1/\mu$. We thus find that GLEMS are characterized
by a peculiar spectrum, with all the
mixedness concentrated in one `decoupled' quadrature. Moreover,
by Eqs.~(\ref{glm}) and (\ref{sympheisweak}) it follows that GLEMS
are minimum--uncertainty mixed Gaussian states.
They are determined by the
standard form correlation coefficients
\begin{eqnarray}
c_{\pm} \!\!&=&\!\! \frac14 \sqrt{\mu_1 \mu_2
\left[-\frac{4}{\mu^2}+\left(1+\frac{1}{\mu^2}-\frac{(\mu_1-\mu_2)^2}{\mu_1^2
\mu_2^2}\right)^2\right]}\nonumber \\
\!\!&\pm&\!\! \frac{1}{4\mu}\sqrt{-4\mu_1\mu_2+\frac{\left[\left(
1+\mu^2\right)\mu_1^2\mu_2^2-\mu^2{\left(\mu_1+\mu_2\right)}^2\right]
^2}{\mu^2\mu_1^3\mu_2^3}} \; . \nonumber \\
\label{glems}
\end{eqnarray}
Quite remarkably, following
the analysis presented in Sec.~\ref{mmix}, it turns out that
the GLEMS at fixed global and marginal purities are also states
of minimal global $p-$entropy for $p<2$,
and of maximal global $p-$entropy for $p>2$.
\par According to the PPT criterion, GLEMS are
separable only if $\mu \le \mu_1 \mu_2 / \sqrt{\mu_1^2 + \mu_2^2
- \mu_1^2 \mu_2^2}$. Therefore, in the range
\begin{equation}
\label{gnslsep}
\frac{\mu_1 \mu_2}{\mu_1 + \mu_2 - \mu_1 \mu_2} < \mu \le
\frac{\mu_1 \mu_2}{\sqrt{\mu_1^2 + \mu_2^2 - \mu_1^2 \mu_2^2}}
\end{equation}
both separable and entangled states can be found. Instead,
the region
\be
\mu>\frac{\mu_1 \mu_2}{\sqrt{\mu_1^2 + \mu_2^2 - \mu_1^2
\mu_2^2}}
\label{sufent}
\ee
can only accomodate \emph{entangled} states.\par
The very narrow region defined by
inequality (\ref{gnslsep}) is thus the only region of coexistence
of both entangled and separable Gaussian mixed states.
We mention that the sufficient condition for entanglement (\ref{sufent}),
first derived in Ref.~\cite{adeser04} has been independently
rediscovered in another recent work \cite{fiurcerf03}.\par

\begin{figure}[t!]
 \includegraphics[width=7.5cm]{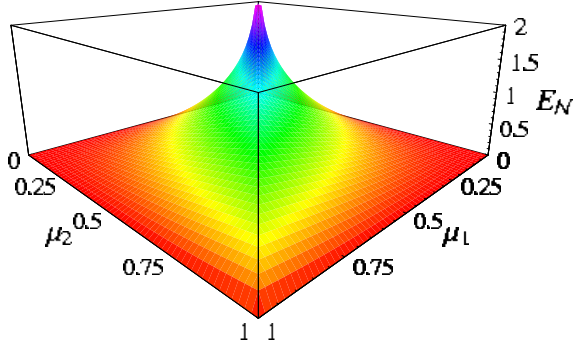}
 \caption{Plot of the maximal logarithmic negativity of
 Gaussian states for fixed marginal purities. The surface represents GMEMMS,
 states that saturate inequality (\ref{upineqmumui}).}
 \label{gmemms}
\end{figure}

\begin{figure*}[t!]
\centering
  \includegraphics[width=12cm]{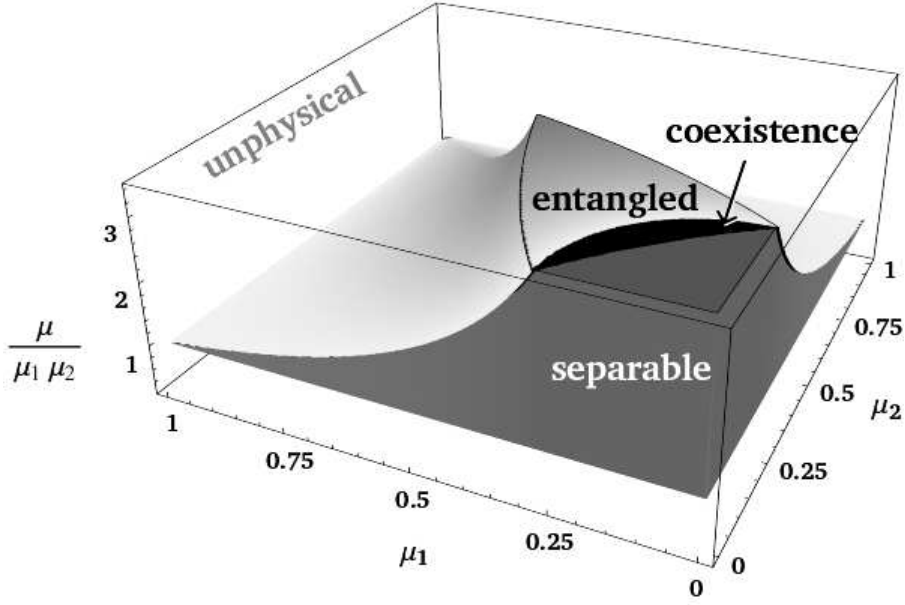}\\
  \caption{
  Summary of entanglement properties of two--mode (nonsymmetric)
  Gaussian states in the space of marginal purities $\mu_{1,2}$
  ($x$- and $y$-axes) and global purity $\mu$. In fact, on the $z$-axis
  we plot the ratio $\mu/\mu_1\mu_2$ to gain a better graphical
  distinction between the various regions. In this space, all physical
  states lay between the horizontal plane $z=1$ representing product states,
  and the upper limiting surface representing GMEMMS. Separable and entangled
  states are well separated except for a narrow region of coexistence (depicted
  in black). Separable states fill the region depicted in dark grey,
  while in the region containing only entangled states we have depicted
  the average logarithmic negativity Eq.~(\ref{average}), growing from white to
  medium grey.
  The mathematical relations defining the boundaries between all these regions
  are collected in Table~\ref{table1}.
  The three-dimensional envelope is cut at $z=3.5$.}
  \label{fig2D}
\end{figure*}

The upper and lower inequalities (\ref{deltabnd}) lead to the
following further constraint between the global and the local
purities
\begin{equation}\label{upineqmumui}
\mu \le \frac{\mu_1 \mu_2}{\mu_1 \mu_2 + \abs{\mu_1-\mu_2}}\,.
\end{equation}
For purities which saturate inequality (\ref{upineqmumui}), one
obtains the states of maximal global purity with given marginal
purities. In such an instance GMEMS and GLEMS coincide and we have
a unique class of entangled states depending only on the marginal
purities $\mu_{1,2}$. They are Gaussian maximally entangled states
for fixed marginals (GMEMMS). In Fig.~\ref{gmemms} the logarithmic
negativity of GMEMMS is plotted as a function of $\mu_{1,2}$,
showing how the maximal entanglement achievable by Gaussian states
rapidly decreases with increasing difference of marginal purities,
in analogy with finite-dimensional systems \cite{adesso03}. For
symmetric states $(\mu_1=\mu_2)$ inequality (\ref{upineqmumui})
reduces to the trivial bound $\mu \le 1$ and GMEMMS reduce to pure
two--mode squeezed vacua.

The previous necessary or sufficient conditions for entanglement
are collected in Table~\ref{table1} and allow a graphical
display of the behavior of the entanglement of mixed Gaussian states
as shown in Fig.~\ref{fig2D}. These relations classify the properties of
separability of all two-mode Gaussian states according to their degree of
global and marginal purities.

\begin{table}[t]
\centering
  \begin{tabular}{|c|c|}
  \hline
  \textsc{Degrees of Purity} & \textsc{Separability}\\ \hline \hline
  $\mu<\mu_1 \mu_2$ & unphysical region \\
  \hline
  $\mu_1 \mu_2 \; \le \; \mu \; \le \; \frac{\mu_1 \mu_2}{\mu_1 + \mu_2 - \mu_1 \mu_2}$ &
  \emph{separable} states
  \\ \hline
  $\frac{\mu_1 \mu_2}{\mu_1 + \mu_2 - \mu_1 \mu_2} < \mu \le
  \frac{\mu_1 \mu_2}{\sqrt{\mu_1^2 + \mu_2^2 - \mu_1^2 \mu_2^2}}$ &
  \emph{coexistence} region \\
   \hline
  $\frac{\mu_1 \mu_2}{\sqrt{\mu_1^2 + \mu_2^2 - \mu_1^2 \mu_2^2}} <
  \mu \le \frac{\mu_1 \mu_2}{\mu_1 \mu_2 + \abs{\mu_1-\mu_2}}$ &
  \emph{entangled} states
  \\ \hline
  $\mu > \frac{\mu_1 \mu_2}{\mu_1 \mu_2 + \abs{\mu_1-\mu_2}}$ &
  unphysical region
  \\ \hline
\end{tabular}
\caption{Classification of two--mode Gaussian states and of their
properties of separability according to their degrees of global
purity $\mu$ and of marginal purities $\mu_1$ and $\mu_2$.}
\label{table1}
 \end{table}

\subsection{Realization of extremally entangled states in experimental settings
\label{exp}}

As we have seen, GMEMS are two-mode squeezed thermal states, whose
general covariance matrix is described by Eqs.~(\ref{stform}) and
(\ref{2mst}). A realistic instance giving rise to such states is
provided by the dissipative evolution of an initially pure
two-mode squeezed vacuum created, \emph{e.g.}, in a non degenerate
parametric down conversion process. Let us denote by
$\gr{\sigma}_r$ the covariance matrix of a two mode squeezed
vacuum with squeezing parameter $r$, derived from
Eqs.~(\ref{2mst}) with $\nu_{\mp}=1$. The interaction of this
initial state with a thermal noise results in the following
dynamical map describing the time evolution of the covariance
matrix $\gr{\sigma}(t)$ \cite{dumodi} \be \gr{\sigma}(t)=\,{\rm
e}^{-\Gamma t}\gr{\sigma}_{r} + (1-\,{\rm e}^{-\Gamma
t})\gr{\sigma}_{n_1,n_2} \; ,\label{realgmems} \ee where $\Gamma$
is the coupling to the noisy reservoir (equal to the inverse of
the damping time) and $\gr{\sigma}_{n_1,n_2}=\oplus_{i=1}^{2} n_i
{\mathbbm 1}_2$ is the covariance matrix of the thermal noise.
The average number of thermal photons $n_i$ given by
\[
n_i=\frac{1}{\exp \left( \hbar \omega_{i}/k_{B}T \right) - 1}
\]
in terms of the frequencies of the modes $\omega_i$ and of the
temperature of the reservoir $T$.
It can be easily verified that the covariance matrix
\eq{realgmems} defines a two-mode thermal squeezed state, generally
nonsymmetric (for $n_1 \neq n_2$). However, notice that the entanglement
of such a state cannot persist indefinetely, becaus after a given time
inequality (\ref{2mstppt}) will be violated and the state will evolve into
a non entangled two-mode squeezed thermal state.
We also notice that the relevant instance of pure loss
($n_1=n_2=0$) allows the realization of symmetric GMEMS.\par
\begin{figure}[b]
\begin{center}
\includegraphics[width=8cm]{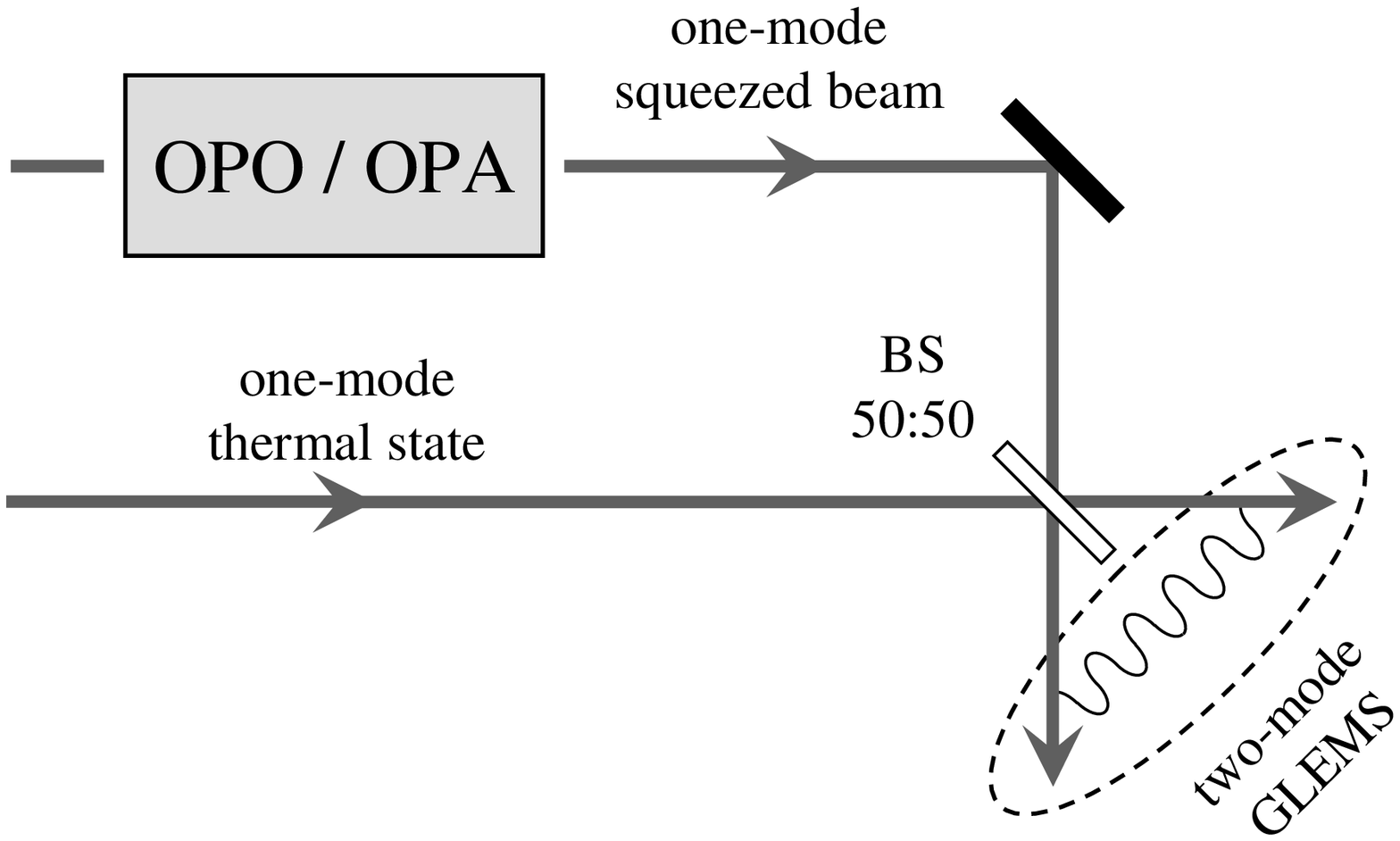}
\caption{Possible scheme for the generation of Gaussian least entangled
mixed states (GLEMS). A single-mode squeezed state (obtained, for example,
by an optical parametric oscillator or amplifier) interferes with a thermal state
through a 50:50 beam splitter. The resulting two-mode state is a minimally entangled
mixed Gaussian state at given global and marginal purities.
\label{glemsetup}}
\end{center}
\end{figure}

Concerning the experimental characterization of minimally entangled states,
one can envisage several explicit
experimental settings for their realiazion. For instance, let
us consider (see Fig.~\ref{glemsetup}) a beam splitter with transmittivity
$\tau=1/2$ (corresponding to a two-mode rotation of angle $\pi/4$
in phase space).
Suppose that a single-mode squeezed state, with covariance matrix
$\gr{\sigma}_{1r}=\,{\rm diag}\,(\,{\rm e}^{2r},\,{\rm e}^{-2r})$
(like, {\em e.g.}, the result of a degenerate parametric down conversion
in a nonlinear crystal), enters in the first input of the beam splitter.
Let the other input be an incoherent thermal state
produced from a source at equilibrium at a temperature $T$. The purity
$\mu$ of such a state can be easily computed in terms of the temperature $T$
and of the frequency of the thermal mode $\omega$
\be
\mu= \frac{\exp \left( \hbar \omega / k_{B} T \right) - 1}{\exp \left( \hbar \omega
/ k_{B} T \right) + 1} \; .
\ee
The state at the output of the beam splitter will be a correlated two-mode
Gaussian state with covariance matrix $\gr{\sigma}_{out}$ that reads
\[
\gr{\sigma}_{out}=\frac{1}{2}\left(
\begin{array}{cccc}
n+k&0&n-k&0\\
0&n+k^{-1}&0&n-k^{-1}\\
n-k&0&n+k&0\\
0&n-k^{-1}&0&n+k^{-1}
\end{array}
\right) \; ,
\]
with $k=\,{\rm e}^{2r}$ and $n=\mu^{-1}$. By immediate inspection,
the symplectic spectrum of this covariance matrix is $\nu_-=1$
and $\nu_+=1/\mu$. Therefore the output state is always a state
with extremal generalized entropy at a given purity
(the state can be seen as the tensor product of
a vacuum and of a thermal state, on which one has applied
symplectic transformation). Moreover, the state is
entangled if
\be
\cosh(2r) > \frac{\mu^2+1}{2\mu} =
\frac{\exp \left( 2\hbar \omega / k_{B} T \right) + 1}{\exp \left(
2\hbar \omega / k_{B} T \right) - 1} \, .
\ee
Tuning the experimental parameters to meet the above condition, indeed
makes the output state of the beam splitter a symmetric GLEMS.
It is interesting to observe that nonsysmmetric GLEMS can be produced
as well by choosing a beam splitter with transmissivity
different from $0.5$.

\section{Extremal entanglement at fixed global and local generalized entropies
\label{genent}}

In this section we introduce a more general characterization of
the entanglement of generic two--mode Gaussian states, by
exploiting the generalized $p-$entropies as measures of global and
marginal mixedness. For ease of comparison we will carry out this
analysis along the same lines followed in the previous Section, by
studying the explicit behavior of the global invariant $\D$,
directly related to the logarithmic negativity $E_\N$ at fixed
global and marginal entropies. This study will clarify the
relation between $\D$ and the generalized entropies $S_p$ and the
ensuing consequences for the entanglement of Gaussian states.

We begin by observing that the
standard form covariance matrix $\gr\sigma$ of a generic two--mode
Gaussian state can be parametrized by the following quantities:
the two marginals $\mu_{1,2}$
(or any other marginal $S_{p_{1,2}}$ because all the local,
single-mode entropies are equivalent for any value of the integer $p$),
the global $p-$entropy $S_p$ (for some chosen value of the
integer $p$), and the global symplectic invariant $\D$. On the other hand,
Eqs.~(\ref{pgen},\ref{pgau},\ref{sympeig}) provide
an explicit expression for any $S_p$ as a function of $\mu$ and $\D$.
Such an expression can be exploited to
study the behaviour of $\D$ as a function of
the global purity $\mu$, at fixed marginals and global $S_p$
(from now on we will omit the explicit reference to fixed marginals).
One has
\bea
\nonumber
\left.\frac{\partial \mu}{\partial
\D}\right\vert_{S_p}=-\frac2{R^2} \left. \frac{\partial R}{\partial
\D}\right\vert_{S_p}=\frac2{R^2} \frac{\left. \partial
S_p / \partial \D \right\vert_{R}}{\left. \partial
S_p / \partial R \right\vert_{\D}}\\
=\frac2{R^2} \frac{N_p(\D,R)}{D_p(\D,R)}\label{dubbino} \; ,
\eea
where we have defined the inverse participation ratio
\be
R\equiv\frac2{\mu} \; ,
\ee
and the remaining quantities $N_p$ and $D_p$ read
\begin{eqnarray}
N_p(\D,R) &=& \left[
(R+2+2\sqrt{\D+R})^{p-1} \right. \nonumber\\
 &-&\left.(R+2-2\sqrt{\D+R})^{p-1} \right] \sqrt{\D-R} \nonumber\\
   &-& \left[
(R-2+2\sqrt{\D-R})^{p-1}\right.\nonumber\\
&-&\left.(R-2-2\sqrt{\D-R})^{p-1} \right]\sqrt{\D+R} \; , \nonumber\\
&& \nonumber \\
D_p(\D,R) &=& \left[(\sqrt{\D+R}+1)
(R+2+2\sqrt{\D+R})^{p-1}\right.\nonumber\\
&+&\left.(\sqrt{\D+R}-1)(R+2-2\sqrt{\D+R})^{p-1}
\right]\nonumber\\
&\cdot&\sqrt{\D-R}\nonumber\\
&-& \left[(\sqrt{\D-R}+1)
(R-2-2\sqrt{\D-R})^{p-1}\right.\nonumber\\
&+&\left.(\sqrt{\D-R}-1)(R-2+2\sqrt{\D-R})^{p-1}
\right]\nonumber\\
&\cdot&\sqrt{\D+R}
\label{npdp} \; .
\end{eqnarray}
Now, it is easily shown that the ratio $N_p(\D,R)/D_p(\D,R)$ is
increasing with increasing $p$ and has a zero at $p=2$ for any
$\D,R$; in particular, its absolute minimum $(-1)$ is reached in
the limit $(\D\rightarrow2,\,R\rightarrow2,\,p\rightarrow1)$. Thus
the derivative \eq{dubbino} is negative for $p<2$, null for $p=2$
(in this case $\D$ and $S_2=1-\mu$ are of course regarded as
independent variables) and positive for $p>2$. This implies that,
for given marginals, keeping fixed any global $S_p$ for $p<2$ the minimum
(maximum) value of $\Delta$ corresponds to the maximum (minimum) value
of the global purity $\mu$. Instead, by keeping fixed any global
$S_p$ for $p>2$ the minimum of $\Delta$ is always attained at the minimum
of the global purity  $\mu$.

This observation allows to determine rather straightforwardly
the states with extremal $\D$. They are
extremally entangled states because,
for fixed global and marginal entropies, the logarithmic negativity
of a state is determined only by the one remaining independent global
symplectic invariant, represented by $\D$ in our choice of parametrization.
If, for the moment being, we neglect the fixed local purities, then the
states with maximal $\D$ are the states with minimal (maximal)
$\mu$ for a given global $S_p$ with $p<2\ (p>2)$ (see Sec.~\ref{mmix} and
Fig.~\ref{svvssl}). As found in Sec.~\ref{mmix}, such states are
minimum uncertainty two--mode states with mixedness
concentrated in one quadrature. We have shown in Sec.~\ref{extre} that
they correspond to Gaussian least entangled mixed states (GLEMS)
whose standard form is given by \eq{glems}.
As can be seen from \eq{glems}, these states are consistent
with any legitimate physical value of the local invariants $\mu_{1,2}$.
We therefore conclude that {\em all} Gaussian states with
{\em maximal} $\D$ for any fixed triple of values of global and marginal
entropies are GLEMS. \par
Viceversa one can
show that {\em all} Gaussian states with {\em minimal} $\D$
for any fixed triple of values of global and marginal
entropies are Gaussian maximally entangled mixed states
(GMEMS). This fact is immediately evident in the symmetric case
because the extremal surface in the  $S_p$ vs. $S_L$ diagrams
is always represented by symmetric two--mode squeezed states (symmetric GMEMS). These
states are characterized by a degenerate symplectic spectrum and
encompass only equal choices of the local invariants:
$\mu_1=\mu_2$. In the nonsymmetric case, the given values of the local entropies
are different, and the extremal value of $\D$ is further constrained
by inequality (\ref{deltabnd})
\be
\D-R\ge\frac{(\mu_1-\mu_2)^2}{\mu_1^2\mu_2^2}\; . \label{56furba}
\ee
From \eq{dubbino} it follows that
\begin{equation}\label{minihorror}
\left.\frac{\partial (\D-R)}{\partial
\D}\right\vert_{S_p,\mu_{1,2}}=1+\frac{N_p(\D,R)}{D_p(\D,R)} \ge
0\,,
\end{equation}
because $N_p(\D,R)/D_p(\D,R)>-1$. Thus, $\D-R$ is
an increasing function of $\D$ at fixed $\mu_{1,2}$ and $S_p$,
and the minimal $\D$ corresponds to the minimum of $\D-R$, which
occurs if inequality (\ref{56furba}) is saturated.
Therefore, also in the nonsymmetric case, the two-mode Gaussian states
with minimal $\D$ at fixed global and marginal entropies are GMEMS.

Summing up, we have shown that GMEMS and GLEMS introduced in the previous section
at fixed global and marginal linear
entropies, are always \emph{extremally} entangled Gaussian
states, whatever triple of generalized global and marginal entropic measures one
chooses to fix.
Maximally and minimally entangled states of CV systems are thus very robust with
respect to the choice of different measures of mixedness. This is at striking
variance with the case of discrete variable systems, where it has been shown
that fixing different measures of mixedness yields different classes of maximally
entangled states \cite{wei03}.

Furthermore, we will now show that the characterization provided
by the generalized entropies leads to some remarkable new insight
on the behavior of the entanglement of CV systems. The crucial
observation is that for a generic $p$, the smallest symplectic
eigenvalue of the partially transposed covariance matrix, at fixed
global and marginal $p-$entropies, is {\em not} in general a
monotone function of $\D$, so that the connection between extremal
$\D$ and extremal entanglement turns out to be, in some cases,
inverted. In particular, while for $p<2$ the GMEMS and GLEMS
surfaces tend to be more separated as $p$ decreases, for $p>2$ the
two classes of extremally entangled states get closer with
increasing $p$ and, within a particular range of global and
marginal entropies, they exchange their role. GMEMS ({\em
i.e.~}states with minimal $\D$) become minimally entangled states
and GLEMS ({\em i.e.~}states with maximal $\D$) become maximally
entangled states. This inversion always occurs for all $p>2$.

To understand this interesting behaviour, let us study the
dependence of the symplectic eigenvalue $\tilde{\nu}_-$ on the
global invariant $\D$ at fixed marginals and at fixed $S_p$ for a
generic $p$. Using Maxwell's relations, we can write
\begin{equation}
\label{maxwell}
\kappa_p\equiv\left.\frac{\partial(2\tilde{\nu}_-^2)}{\partial
\D}\right\vert_{S_p}=\left.\frac{\partial(2\tilde{\nu}_-^2)}{\partial
\D}\right\vert_{R}-\left.\frac{\partial(2\tilde{\nu}_-^2)}{\partial
R}\right\vert_{\D} \cdot \frac{\left. \partial S_p / \partial
\D \right\vert_{R}}{\left. \partial S_p / \partial
R \right\vert_{\D}}\,.
\end{equation}
Clearly, for $\kappa_p>0$ GMEMS and GLEMS retain their usual
interpretation, whereas for $\kappa_p<0$ they exchange their role.
On the \emph{node} $\kappa_p=0$ GMEMS and GLEMS share the same entanglement,
\emph{i.e.}~the entanglement of all Gaussian states at $\kappa_p=0$
is fully determined by the global and marginal $p-$entropies alone,
and does not depend any more on $\D$. Such nodes also exist in the case
$p \le 2$ in two limiting instances: in the special case of GMEMMS (states with maximal
global purity at fixed marginals) and in the limit of zero marginal purities.
We will now show that, besides these two limiting behaviors, a nontrivial node appears
for all $p>2$, implying that on the two sides of the node GMEMS and
GLEMS indeed exhibit opposite behaviors. Because of \eq{dubbino}, $\kappa_p$ can
be written in the following form
\begin{equation}
\label{horror}
\kappa_p=\kappa_2- \frac{R}{\sqrt{\tilde{\D}^2-R^2}} \,
\frac{N_p(\D,R)}{D_p(\D,R)} \; ,
\end{equation}
with $N_p$ and $D_p$ defined by \eq{npdp} and
\begin{eqnarray*}
\tilde{\D} &=& -\D+\frac2{\mu_1^2} + \frac2{\mu_2^2} \; , \\
\kappa_2 &=&
-1+\frac{\tilde{\Delta}}{\sqrt{\tilde{\D}^2-R^2}} \; ,
\end{eqnarray*}

\begin{figure}[tb!]
\subfigure[] {\includegraphics[width=6cm]{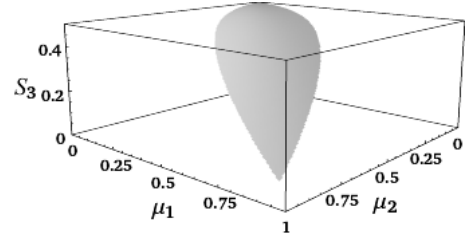}} \subfigure[]
{\includegraphics[width=6cm]{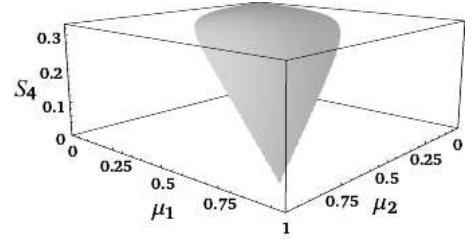}} \caption{Plot of the nodal
surface which solves the equation $\kappa_p=0$ with $\kappa_p$
defined by \eq{horror}, for (a) $p=3$ and (b) $p=4$. The
entanglement of Gaussian states that lie on the leaf--shaped
surfaces is fully quantified in terms of the marginal purities and
the global generalized entropy (a) $S_3$ or (b) $S_4$. The equations
of the surfaces in the space
$\mathcal{E}_p\equiv\{\mu_1,\mu_2,S_p\}$ are given by
Eqs.~(\ref{spleaf}--\ref{spleaf4}).} \label{figleaf}
\end{figure}

The quantity $\kappa_p$ in \eq{horror} is a function of
$p,\,R,\,\D,\,$ and of the marginals; since we are looking for the
node (where the entanglement is independent of $\D$), we can
investigate the existence of a nontrivial solution to the equation
$\kappa_p=0$ fixing any value of $\D$. Let us choose $\D=1+R^2/4$
that saturates the Heisenberg uncertainty relation and is
satisfied by GLEMS. With this position, \eq{horror} becomes
\begin{eqnarray}
\label{horrorl}
\kappa_p(\mu_1,\mu_2,R)&=&\kappa_2(\mu_1,\mu_2,R)\nonumber\\
&-&\frac{R}{\sqrt{\left(
\frac2{\mu_1^2}+\frac2{\mu_2^2}-\frac{R^2}4-1\right)^2-R^2}}\, f_p(R)
\, .
\nonumber \\
\end{eqnarray}
\noindent The existence of the node depends then on the behavior
of the function
\begin{equation}
\label{fp}
f_p(R) \equiv \frac{2 \left[ (R+2)^{p-2}-(R-2)^{p-2} \right]}
{(R+4)(R+2)^{p-2}-(R-4)(R-2)^{p-2}} \; .
\end{equation}
In fact, as we have already pointed out, $\kappa_2$ is always
positive, while the function $f_p(R)$ is an increasing function of
$p$ and, in particular, it is negative for $p<2$, null for $p=2$
and positive for $p>2$, reaching its asymptote $2/(R+4)$ in the
limit $p \rightarrow \infty$. This entails that, for $p\le2$,
$\kappa_p$ is always positive, which in turn implies that GMEMS
and GLEMS are respectively maximally and minimally entangled
two--mode states with fixed marginal and global $p-$entropies in
the range $p\le2$ (including both von Neumann and linear
entropies). On the other hand, for any $p>2$ one node can be found
solving the equation $\kappa_p(\mu_{1},\mu_{2},2/\mu)=0$. The
solutions to this equation can be found analytically for low $p$
and numerically for any $p$. They form a continuum in the space
$\{\mu_1,\mu_2,\mu\}$ which can be expressed as a surface of
general equation $\mu=\mu^\kappa_p(\mu_1,\mu_2)$. Since the fixed
variable is $S_p$ and not $\mu$ it is convenient to rewrite the
equation of this surface in the space
$\mathcal{E}_p\equiv\{\mu_1,\mu_2,S_p\}$, keeping in mind the
relation~(\ref{spglems}), holding for GLEMS, between $\mu$ and
$S_p$. In this way the nodal surface $(\kappa_p=0)$ can be written
in the form
\begin{equation}
\label{spleaf}
S_p=S_p^\kappa(\mu_1,\mu_2) \equiv
\frac{1 - g_p \left[ (\mu_p^\kappa(\mu_1,\mu_2))^{-1}\right]}{p-1} \; .
\end{equation}
The entanglement of all Gaussian states whose entropies lie on the
surface $S_p^\kappa(\mu_1,\mu_2)$ is \emph{completely}
determinated by the knowledge of $\mu_1$, $\mu_2$ and $S_p$. The
explicit expression of the function $\mu^\kappa_p(\mu_1,\mu_2)$
depends on $p$ but, being the global purity of physical
states, is constrained by the inequality
$$
\mu_1 \mu_2 \, \le \, \mu^\kappa_p (\mu_1,\mu_2) \, \le \,
\frac{\mu_1 \mu_2}{\mu_1 \mu_2 + \abs{\mu_1-\mu_2}} \; .
$$
The nodal surface of \eq{spleaf} constitutes
a `leaf', with base at the point
$\mu^\kappa_p(0,0)=0$ and tip at the point
$\mu^\kappa_p(\sqrt3/2,\sqrt3/2)=1$, for any $p>2$; such a leaf
becomes larger and flatter with increasing $p$ (see
Fig.~\ref{figleaf}).
For $p>2$, the function $f_p(R)$ defined by \eq{fp} is negative
but decreasing with increasing $R$, that is with decreasing $\mu$.
This means that, in the space of entropies $\mathcal{E}_p$, above
the leaf $(S_p>S_p^\kappa)$ GMEMS (GLEMS) are still maximally
(minimally) entangled states for fixed global and marginal
generalized entropies, while below the leaf they are
\emph{inverted}. Notice also that for $\mu_{1,2}>\sqrt3/2$ no node
and so no inversion can occur for any $p$. Each point on the
leaf--shaped surface of \eq{spleaf} corresponds to an entire class
of infinitely many two--mode Gaussian states (including GMEMS and
GLEMS) with the same marginals and the same global
$S_p=S_p^\kappa(\mu_1,\mu_2)$, which are all equally entangled,
since their logarithmic negativity is completely determined by
$\mu_1,\mu_2$ and $S_p$. For the sake of clarity we provide the
explicit expressions of $\mu_p^\kappa(\mu_1,\mu_2)$, as plotted in
Fig.~\ref{figleaf} for the cases (a) $p=3$, and (b) $p=4$.
\begin{eqnarray}
  \mu_3^\kappa(\mu_1,\mu_2) &=&
  \left(\frac{6}{\frac{3}{\mu_1^2}+\frac{3}{\mu_1^2}-2}
  \right)^{\frac12},
  \label{spleaf3}\\
&& \nonumber \\
\mu_4^\kappa(\mu_1,\mu_2) &=& \sqrt{3}\,\mu_1 \mu_2\,\,\,
  \bigg/\,\,\,  \bigg(\mu_1^2+\mu_2^2-2\mu_1^2\mu_2^2+ \nonumber\\
  & & \sqrt{\left(\mu_1^2+\mu_2^2\right)\left(\mu_1^2+\mu_2^2-\mu_1^2 \mu_2^2\right)
  +\mu_1^4 \mu_2^4} \bigg)^{\frac12}.\nonumber \\
\label{spleaf4}
\end{eqnarray}

\begin{figure*}[t!]
\centering
 \hspace{2mm}
 \subfigure[\label{fig2D1}]
{\includegraphics[width=6.7cm]{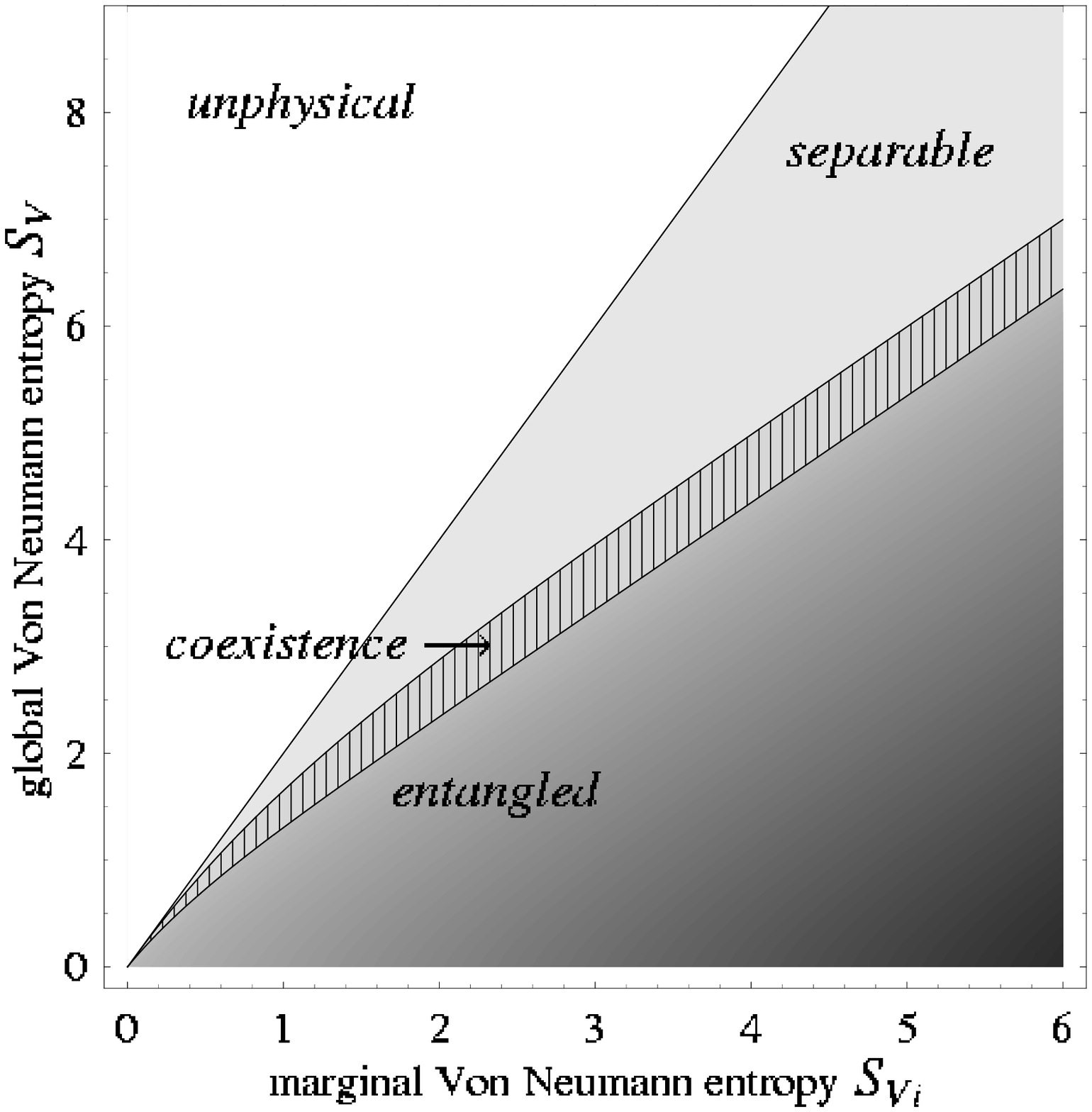}}
 \hspace{7mm}
\subfigure[\label{fig2D2}]
{\includegraphics[width=6.9cm]{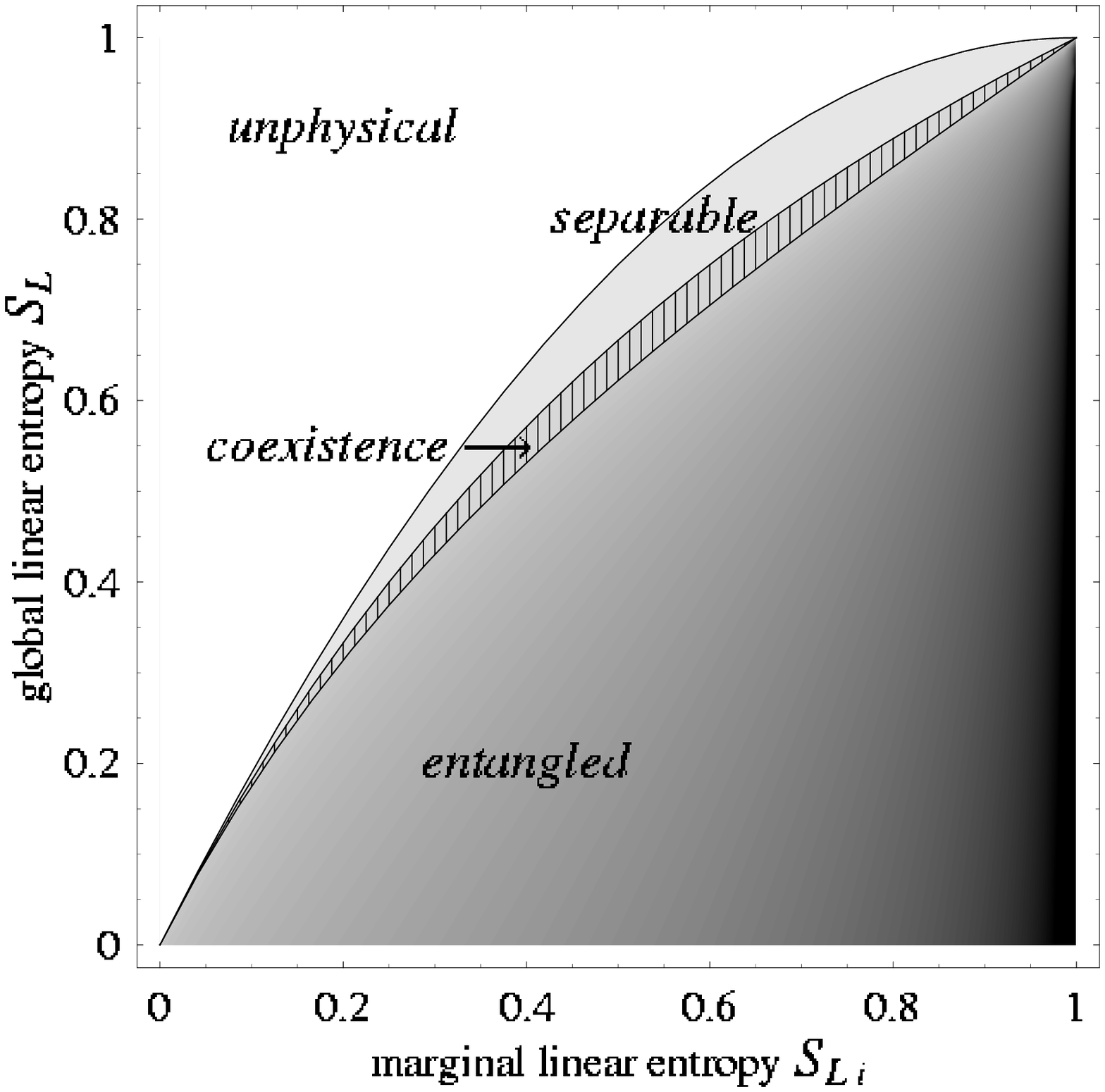}}
\subfigure[\label{fig2D3}]
{\includegraphics[width=7.0cm]{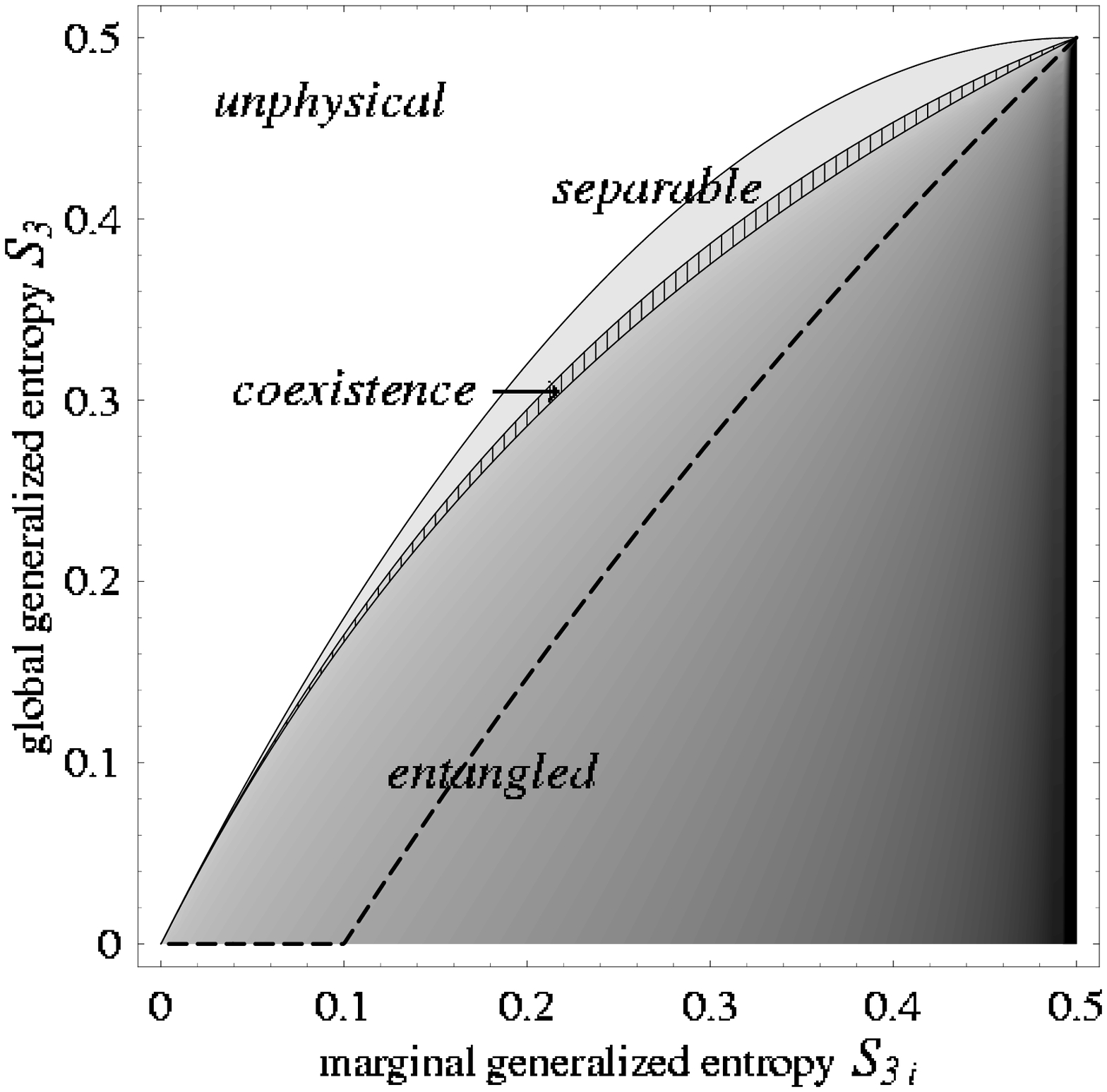}}
 \hspace{5mm}
\subfigure[\label{fig2D4}]
{\includegraphics[width=7.1cm]{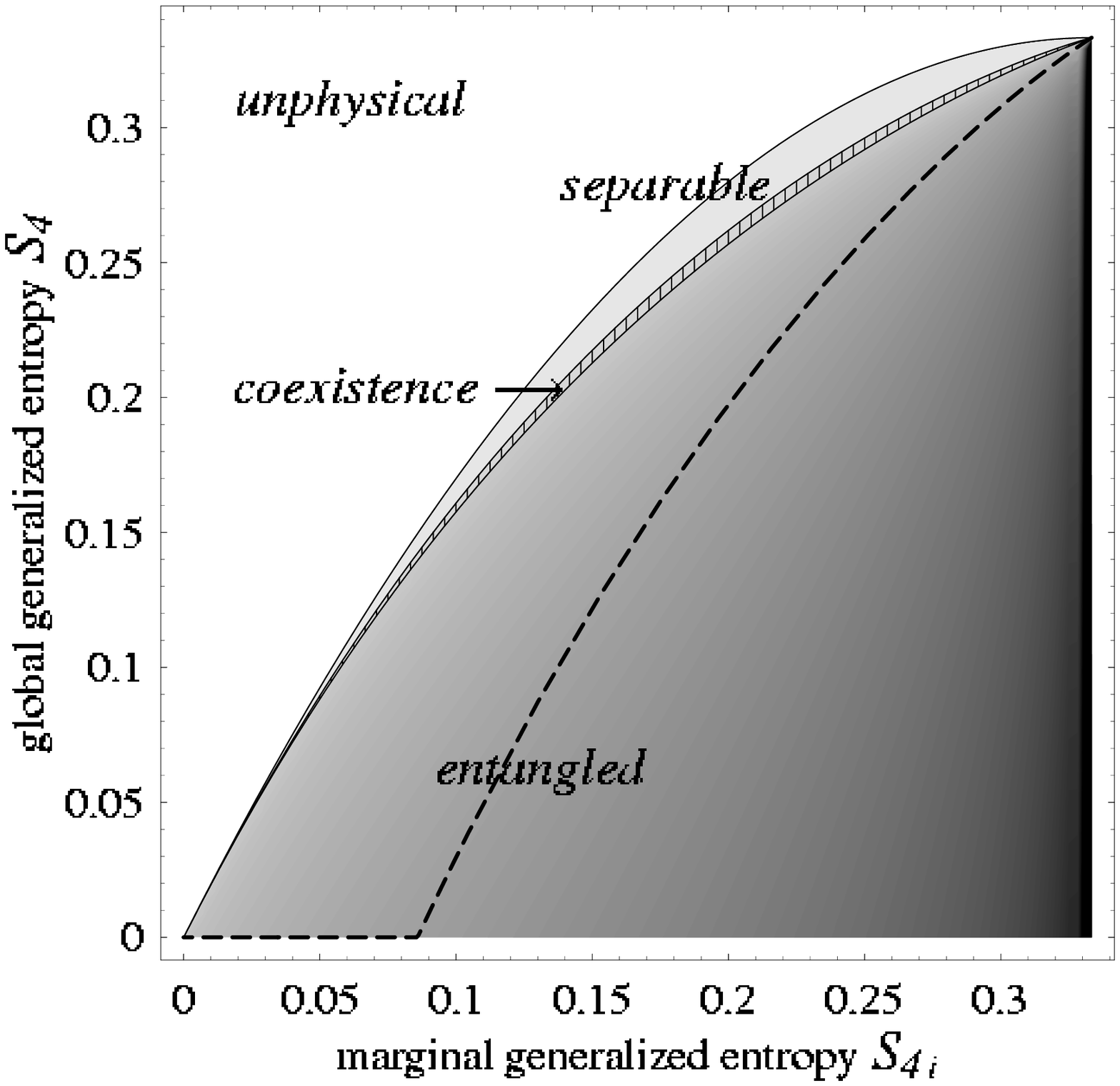}}
  \caption{Summary of the entanglement properties for symmetric Gaussian
  states at fixed global and marginal generalized $p-$entropies, for
  (a) $p=1$ (von Neumann entropies), (b) $p=2$ (linear entropies), (c) $p=3$,
  and (d) $p=4$.  In the entangled region, the average logarithmic
  negativity $\bar{E}_{\N}(S_{p_i},S_p)$ \eq{average} is depicted, growing from gray to
  black. For $p>2$ an additional dashed curve is plotted;
  it represents the nodal line of inversion. Along it the entanglement is fully
  determined by the knowledge of the global and marginal generalized entropies
  $S_{p_i},S_p$, and GMEMS and GLEMS are equally entangled. On the left side of the nodal line GMEMS
  (GLEMS) are maximally (minimally) entangled Gaussian states at fixed $S_{p_i},S_p$.
  On the right side of the nodal line they are inverted: GMEMS (GLEMS) are minimally
  (maximally) entangled states. Also notice how the dashed region of
  coexistence becomes narrower with increasing $p$. The equations
  of all boundary curves can be found in \eq{entsump}.}
\label{fig2Dm}
\end{figure*}

Apart from the relevant `inversion' feature shown by $p-$entropies
for $p>2$, the possibility of an accurate characterization of CV
entanglement based on global and marginal entropic measures still
holds in the general case for any $p$. In particular, the set of
all Gaussian states can be again divided, in the space of global
and marginal $S_p$'s, into three main areas: separable, entangled
and coexistence region. It can be thus very interesting to
investigate how the different entropic measures chosen to quantify
the degree of global mixedness (all marginal measures are
equivalent) behave in classifying the separability properties of
Gaussian states.  Fig.~\ref{fig2Dm} provides a numerical
comparison of the different characterizations of entanglement
obtained by the use of different $p-$entropies, with $p$ ranging
from $1$ to $4$, for symmetric Gaussian states
($S_{p_1}=S_{p_2}\equiv S_{p_i}$). The last restriction has been
imposed just for ease of graphical display. The following
considerations, based on the exact numerical solutions of the
transcendental conditions, will take into account nonsymmetric
states as well.

The mathematical relations expressing the boundaries between the
different regions in Fig.~\ref{fig2Dm} are easily obtained for any
$p$ by starting from the relations holding for $p=2$ (see
Table~\ref{table1}) and by evaluating the corresponding
$S_p(\mu_{1,2})$ for each $\mu(\mu_{1,2})$. For any physical
symmetric state such a calculation yields
\begin{eqnarray}
 0  \le   (p-1) S_p &\!\!<\!\!&
1-g_p\left(\frac{\sqrt{2-\mu_i^2}}{\mu_i}\right)\nonumber\\
  &\Rightarrow& \textrm{entangled,} \nonumber\\
&& \nonumber \\
  1-g_p\left(\frac{\sqrt{2-\mu_i^2}}{\mu_i}\right)
\le  (p-1) S_p &\!\!<\!\!&
1-g_p^2\left(\sqrt{\frac{2-\mu_i}{\mu_i}}\right)\nonumber\\
&\Rightarrow& \textrm{coexistence,} \nonumber\\
&& \nonumber \\
1-g_p^2\left(\sqrt{\frac{2-\mu_i}{\mu_i}}\right)  \le  (p-1)
S_p &\!\! \le \!\!& 1-g_p^2\left(\frac1{\mu_i^2}\right)\nonumber\\
  &\Rightarrow& \textrm{separable.}
  \label{entsump}
\end{eqnarray}
Equations~(\ref{entsump}) were obtained exploiting the
multiplicativity of $p-$norms on product states and using
\eq{spmin} for the lower boundary of the coexistence region (which
represents GLEMS becoming entangled) and \eq{spmax} for the upper
one (which expresses GMEMS becoming separable). Let us mention
also that the relation between any local entropic measure
$S_{p_i}$ and the local purity $\mu_i$ is obtained directly from
\eq{pgau} and reads
\begin{equation}\label{spimui}
S_{p_i}=\frac{1-g_p(1/\mu_i)}{p-1}\,.
\end{equation}

We notice {\em prima facie} that, with increasing $p$, the
entanglement is more sharply qualified in terms of the global and
marginal $p-$entropies. In fact the region of coexistence between
separable and entangled states becomes narrower with higher $p$.
Thus, somehow paradoxically, with increasing $p$ the entropy $S_p$
provides less information about a quantum state, but at the same
time it yields a more accurate characterization and quantification
of its entanglement. In the limit $p \rightarrow \infty$ all the
physical states collapse to one point at the origin of the axes in
the space of generalized entropies, due to the fact that the
measure $S_\infty$ is identically zero.
\begin{figure*}[t!]
\subfigure[\label{fig3D1}]
{\includegraphics[width=6.5cm]{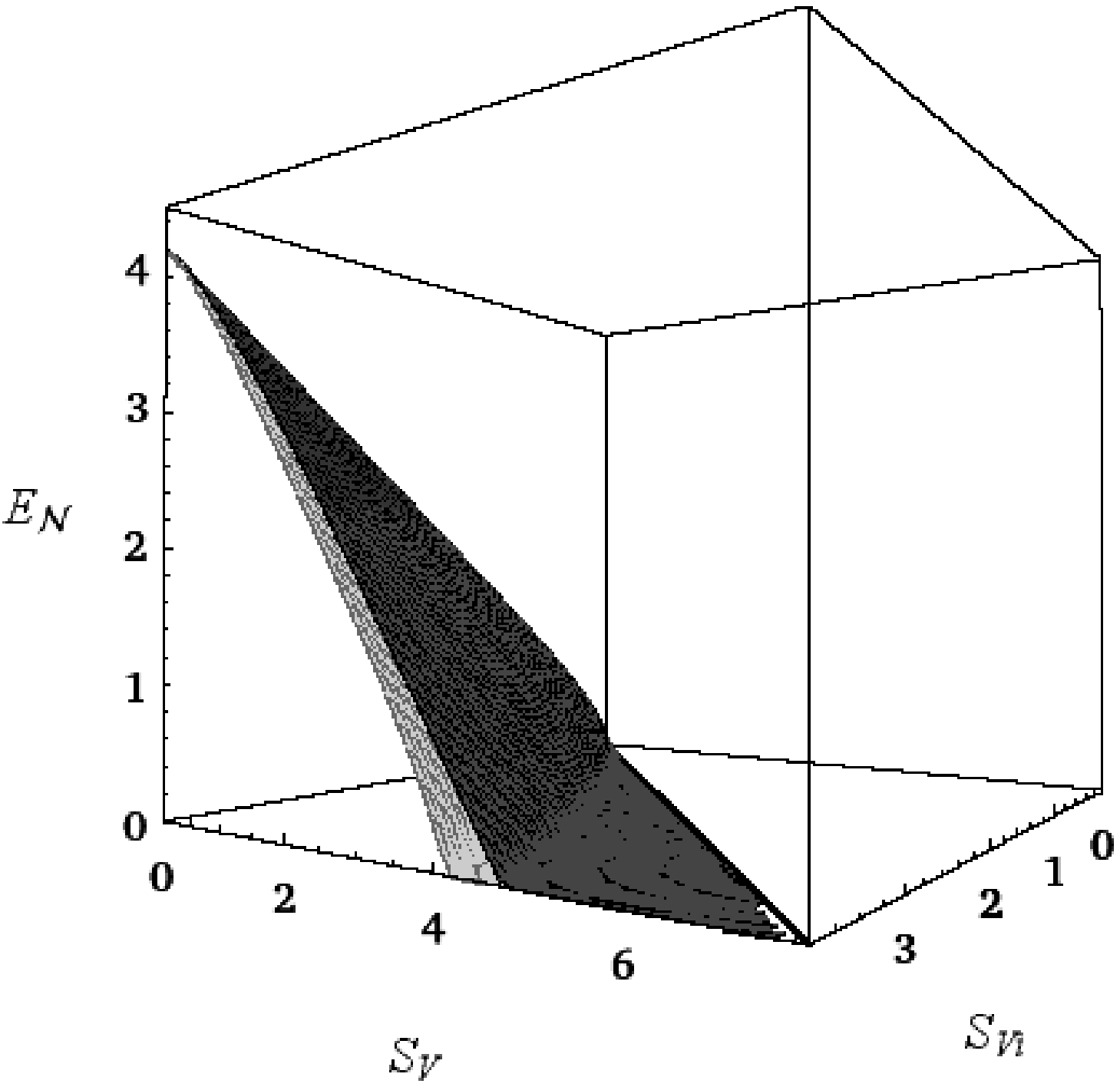}} \hspace{5mm}
\subfigure[\label{fig3D2}]
{\includegraphics[width=6.5cm]{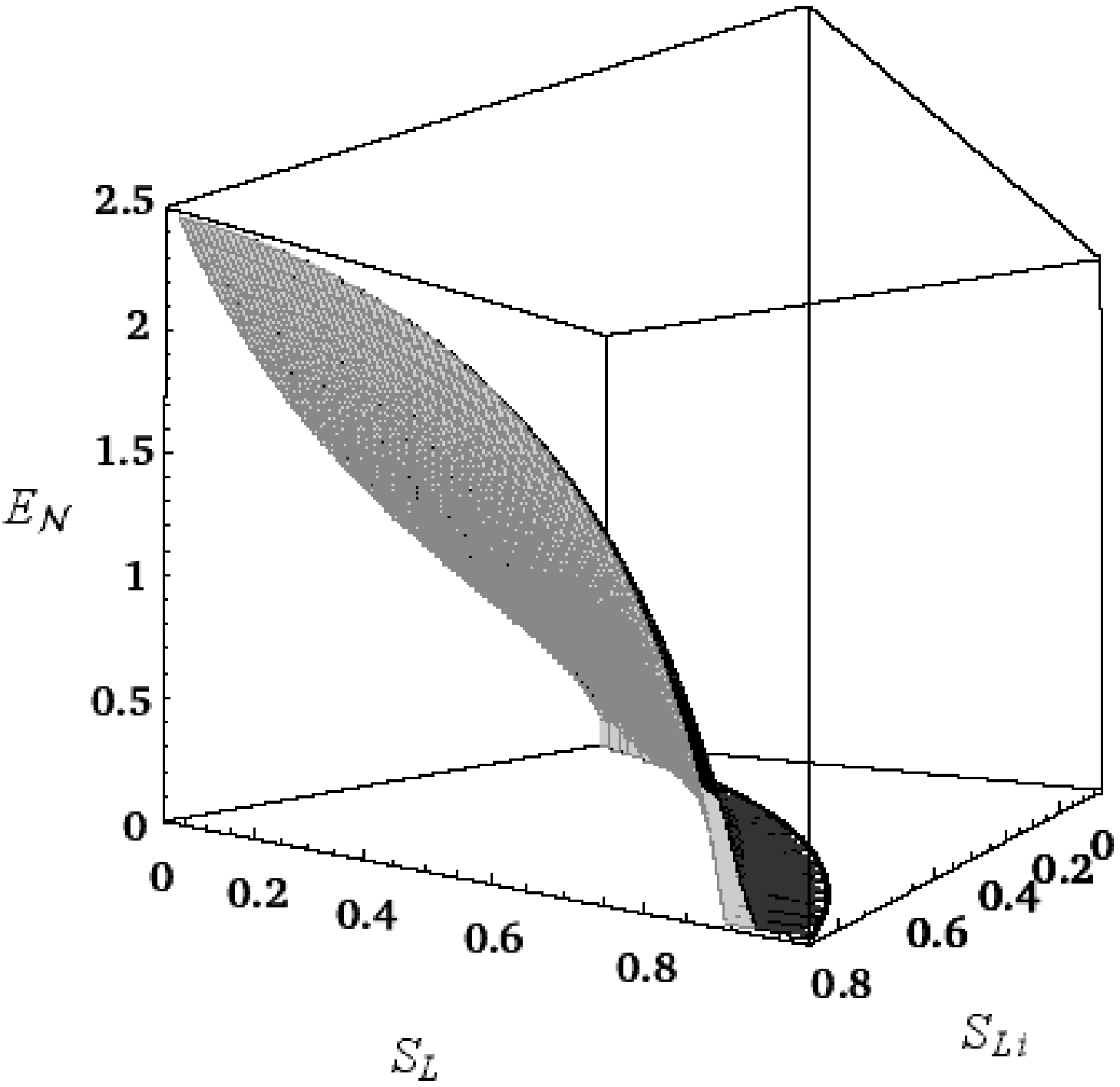}}
\subfigure[\label{fig3D3}]
{\includegraphics[width=6.5cm]{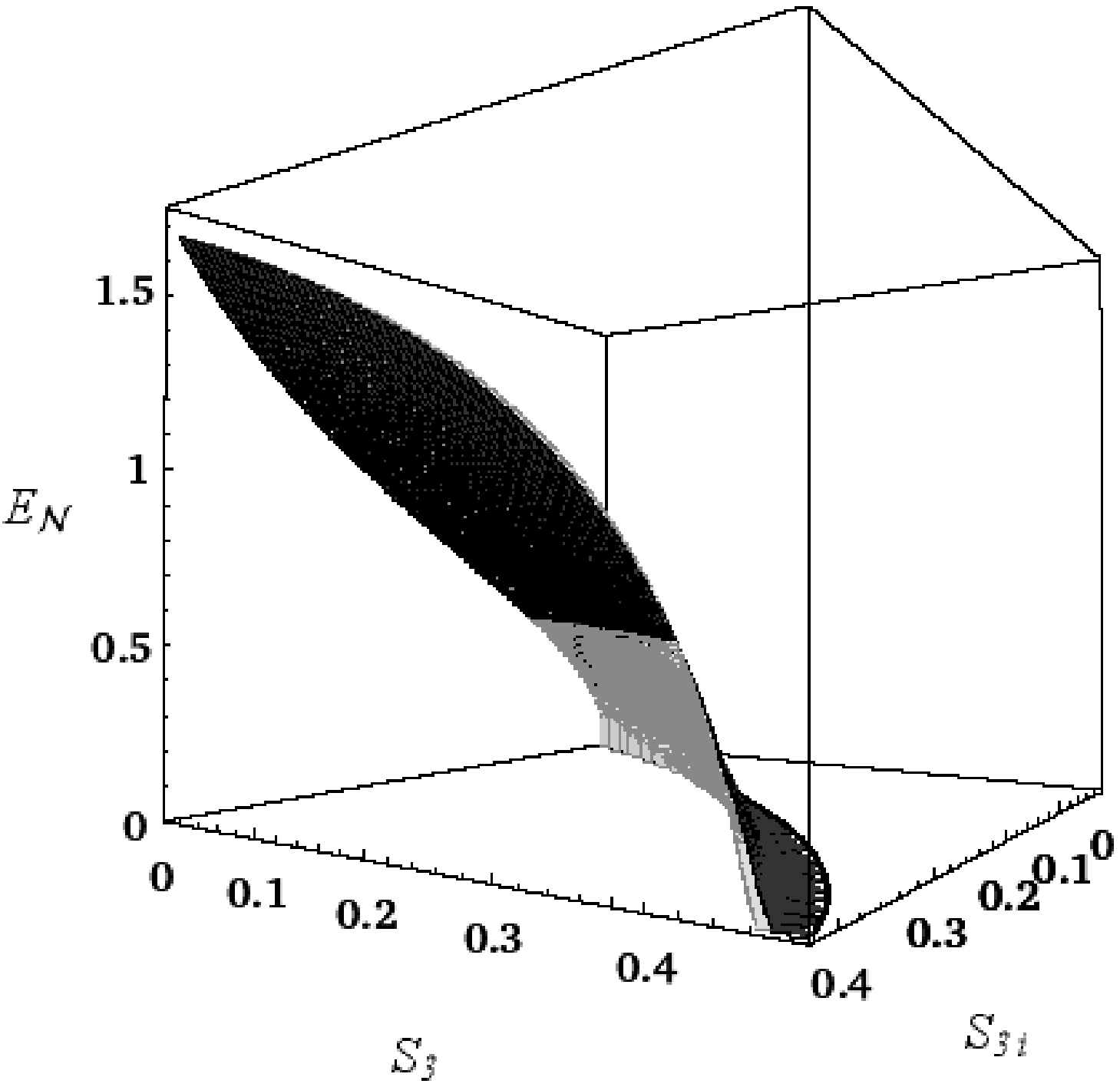}} \hspace{5mm}
\subfigure[\label{fig3D4}]
{\includegraphics[width=6.5cm]{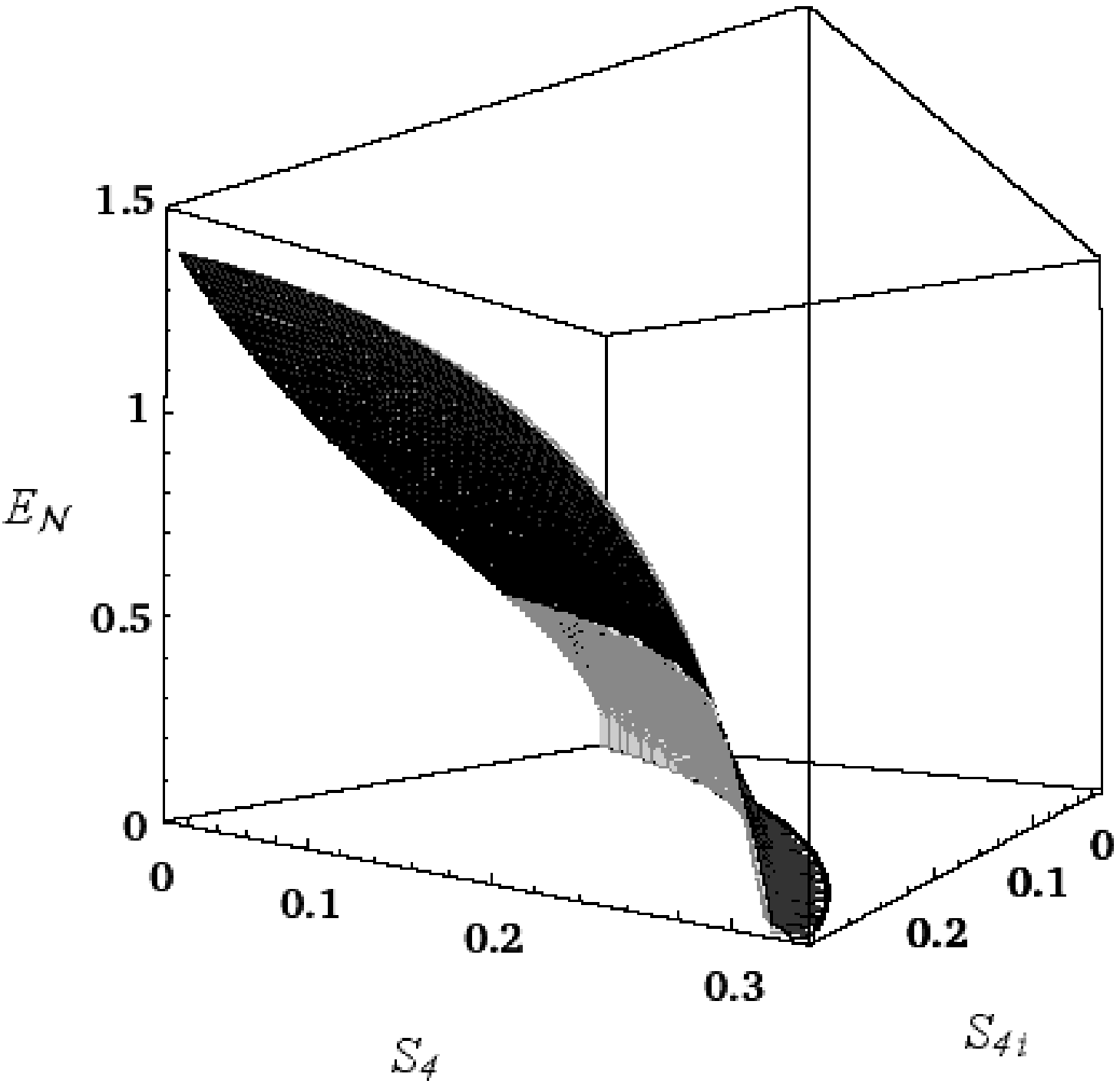}}
  \caption{Upper and lower bounds on the logarithmic
  negativity of symmetric Gaussian states
  as functions of the global and marginal generalized $p-$entropies, for
  (a) $p=1$ (von Neumann entropies), (b) $p=2$ (linear entropies), (c) $p=3$,
  and (d) $p=4$. The black (gray) surface represents GMEMS
  (GLEMS). Notice that for $p>2$ GMEMS and GLEMS surfaces intersect
along the inversion line (meaning they are equally entangled along
that line), and beyond it they interchange their role. The
equations of the inversion lines are obtained from
Eqs.~(\ref{spleaf}--\ref{spleaf4}), with the position
$S_{p_1}=S_{p_2}\equiv S_{p_i}$.}
  \label{fig3Dm}
\end{figure*}

\section{Quantifying Entanglement: the average logarithmic negativity\label{aln}}

We have extensively shown that knowledge of the global and
marginal generalized $p-$entropies accurately characterizes the
entanglement of Gaussian states, providing strong sufficient or
necessary conditions. The present analysis naturally leads us to
propose an actual \emph{quantification} of entanglement, based
exclusively on marginal and global entropic measures, enriching
and generalizing the approach introduced in Ref.~\cite{adeser04}.

Outside the separable region, we can formally define the maximal
entanglement $E_{{\N}max}(S_{p_{1,2}},S_p)$ as the logarithmic
negativity attained by GMEMS (or GLEMS, below the inversion nodal surface
for $p>2$, see Fig.~\ref{figleaf}).
In a similar way, in the entangled region GLEMS (or
GMEMS, below the inversion nodal surface for $p>2$)
achieve the minimal logarithmic negativity
$E_{{\N}max}(S_{p_{1,2}},S_p)$. The
explicit analytical expressions of these quantities are
unavailable for any $p\neq2$ due to the transcendence of the
conditions relating $S_p$ to the symplectic eigenvalues. The
surfaces of maximal and minimal entanglement in the space of the
global and local $S_p$ are plotted in Fig.~\ref{fig3Dm} for
symmetric states. In the plane $S_p=0$ the upper and lower
bounds correctly coincide, since for pure states the entanglement
is completely quantified by the marginal entropy. For mixed states
this is not the case but, as the plot shows, knowledge of the
global and marginal entropies strictly bounds the entanglement both
from above and from below. For $p>2$, we notice how GMEMS and
GLEMS exchange their role beyond a specific curve in the space of
$S_p$'s. The equation of this nodal curve is obtained from the
general leaf--shaped nodal surfaces of
Eqs.~(\ref{spleaf}--\ref{spleaf4}), by imposing the symmetry
constraint ($S_{p_1}=S_{p_2}\equiv S_{p_i}$). We notice again how
the $S_p$'s with higher $p$ provide a better characterization of
the entanglement, even quantitatively.
In fact, the gap between the two extremally
entangled surfaces in the $S_p$'s space becomes smaller with
higher $p$. Of course the gap is exactly zero all along the nodal
line of inversion for $p>2$.

\begin{figure*}[t!]
\subfigure[\label{figerror1}]
{\includegraphics[width=5.5cm]{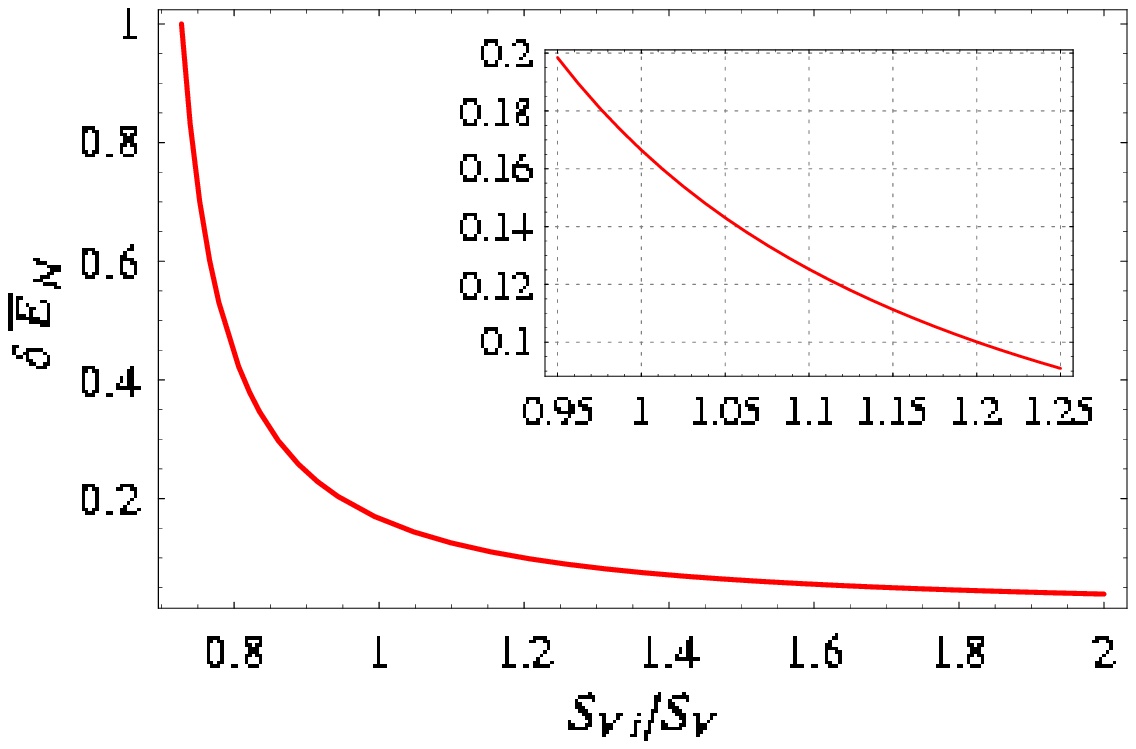}} \hspace{7mm}
\subfigure[\label{figerror2}]
{\includegraphics[width=5.5cm]{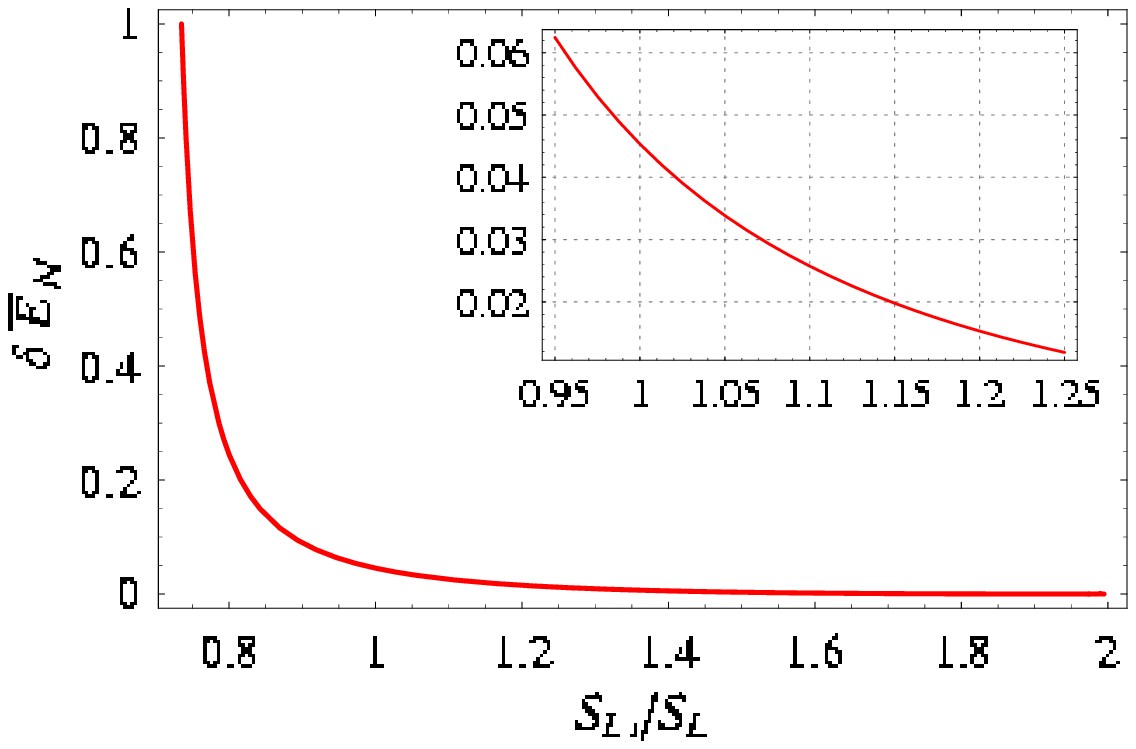}}\\
\subfigure[\label{figerror3}]
{\includegraphics[width=5.5cm]{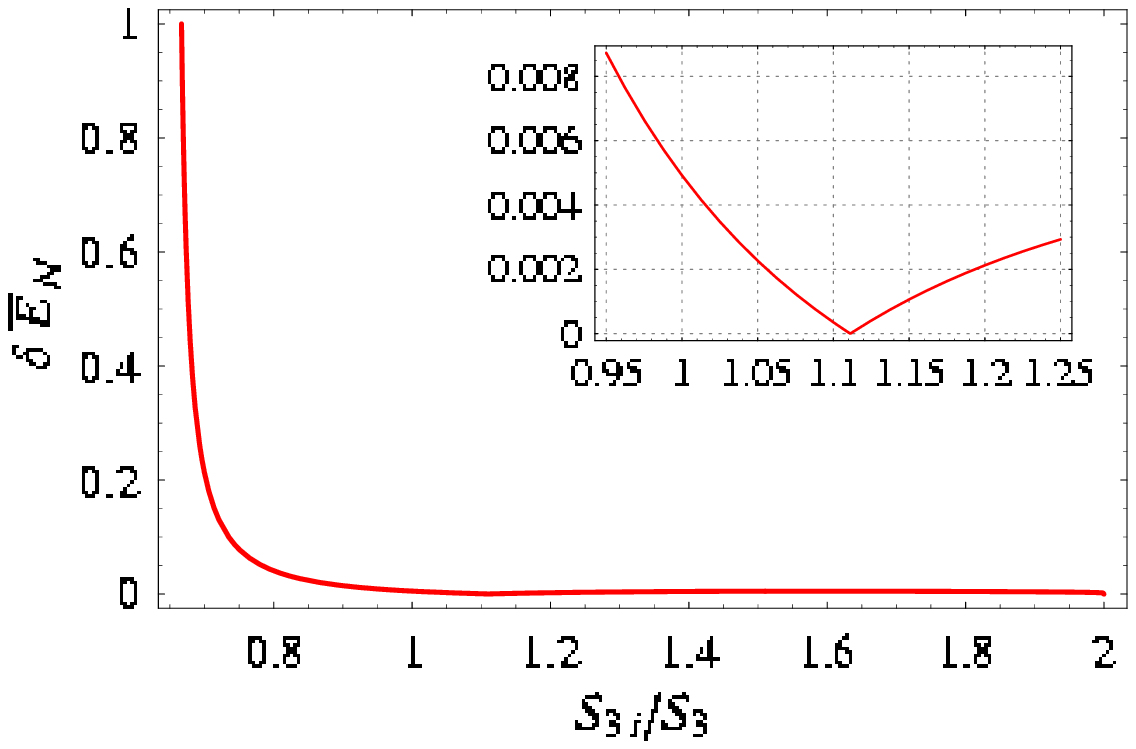}} \hspace{7mm}
\subfigure[\label{figerror4}]
{\includegraphics[width=5.5cm]{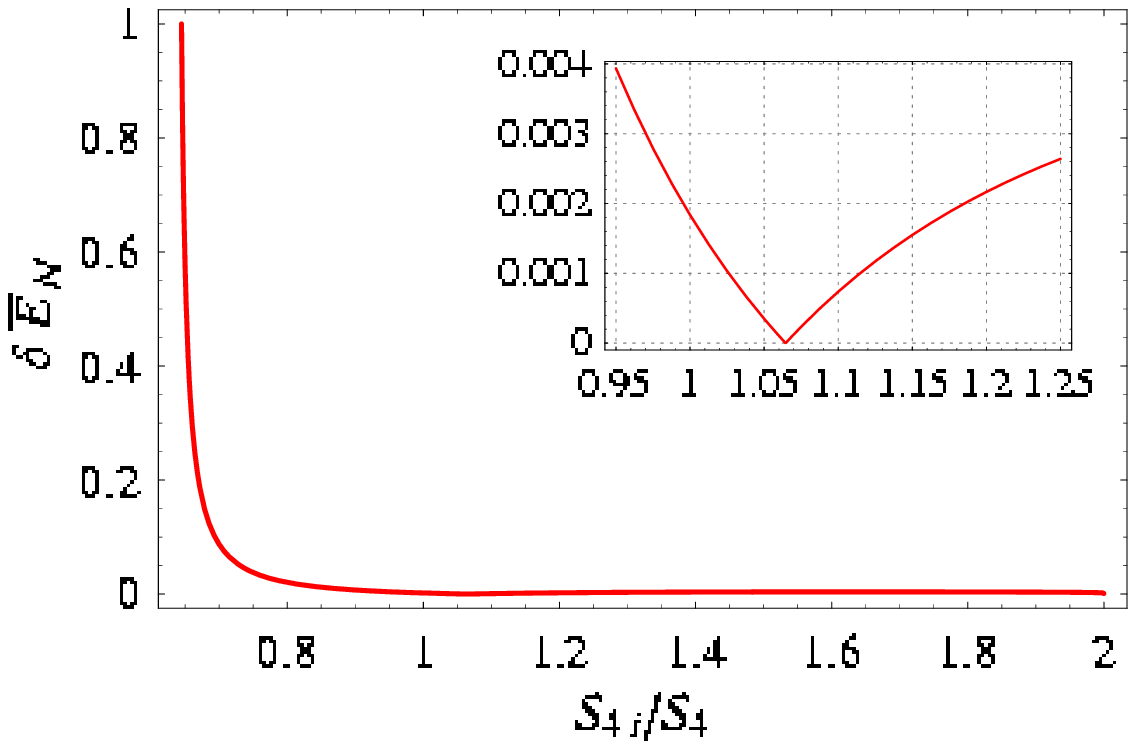}} \caption{The relative
error $\delta \bar{E}_{\N}$ Eq.~(\ref{deltaen}) on the average
logarithmic negativity as a function of the ratio $S_{p_i} /S_p$,
for (a) $p=1$, (b) $p=2$, (c) $p=3$, (d) $p=4$, plotted at (a)
$S_V=1$, (b) $S_L=1/2$, (c) $S_3=1/4$, (d) $S_4=1/6$. Notice how, in
general, the error decays exponentially, and in particular faster
with increasing $p$. For $p>2$, notice how the error reaches zero on
the inversion node (see the insets), then grows and reaches a local
maximum before going back to zero asymptotically.}
  \label{figerrorm}
\end{figure*}

We will now introduce a particularly convenient
quantitative estimate of the entanglement
based only on the knowledge of the
global and marginal entropies.
Let us define
the ``average logarithmic negativity'' $\bar{E}_{\N}$ as
\be
\bar{E}_{\N} (S_{p_{1,2}},S_p) \equiv
\frac{E_{{\N}max}(S_{p_{1,2}},S_p)+E_{{\N}min}(S_{p_{1,2}},S_p)}{2}
\; . \label{average}
\ee
We will now show that this quantity,
fully determined by the global and marginal entropies, provides a
reliable quantification of entanglement (logarithmic negativity)
for two--mode Gaussian states. To this aim, we define the
relative error $\delta \bar{E}_{\N}$ on $\bar{E}_{\N}$ as
\begin{equation}
\label{deltaen}
\delta \bar{E}_{\N} (S_{p_{1,2}},S_p) \equiv
\frac{E_{Nmax}(S_{p_{1,2}},S_p)
-E_{Nmin}(S_{p_{1,2}},S_p)}{E_{Nmax}(S_{p_{1,2}},S_p)
+E_{Nmin}(S_{p_{1,2}},S_p)}\,.
\end{equation}
As Fig.~\ref{figerrorm} shows, this error decreases
\emph{exponentially} both with decreasing global entropy and
increasing marginal entropies, that is with increasing
entanglement. In general the relative error $\delta \bar{E}_{\N}$
is `small' for sufficiently entangled states; we will present
more precise numerical considerations in the subcase $p=2$.
Notice that the decaying rate of the relative error is faster with
increasing $p$: the average logarithmic negativity turns out
to be a better estimate of entanglement with increasing $p$. For
$p>2$, $\delta \bar{E}_{\N}$ is exactly zero on the inversion node,
then it becomes finite again and, after reaching a local maximum,
it goes asymptotically to zero (see the insets of Fig.~\ref{figerrorm}).

All the above considerations, obtained by an exact numerical analysis,
show that the average logarithmic negativity $\bar{E}_{\N}$ at
fixed global and marginal $p-$entropies is a very good
estimate of entanglement in CV systems, whose reliability
improves with increasing
entanglement and, surprisingly, with increasing order $p$ of the
entropic measures.

\subsection{Direct estimate of the entanglement}

In the present general framework, a peculiar
role is played by the case $p=2$, {\em i.e.} by the linear
entropy $S_L$ (or, equivalently, the purity $\mu$).
The previous general analysis on the whole
range of generalized entropies $S_p$, has remarkably stressed the
privileged theoretical role of the instance $p=2$, which
discriminates between the region in which extremally entangled
states are unambiguously characterized and the region in which they
can exchange their roles. Moreover, the graphical analysis
shows that, in the region where no inversion takes
place $(p \le 2)$, fixing the global $S_2=1-\mu$ yields the most
stringent constraints on the logarithmic negativity of the states
(see Figs.~\ref{fig2Dm}, \ref{fig3Dm}, \ref{figerrorm}). Notice
that such constraints, involving no transcendental functions for
$p=2$, can be easily handled analytically.
A crucial experimental consideration strengthens
these theoretical and practical reasons to
privilege the role of $S_2$.
In fact, $S_2$ can indeed, assuming some prior
knowledge about the state (essentially, its Gaussian character),
be measured through conceivable direct methods, in particular by
means of single--photon detection schemes \cite{fiurcerf03} or of
the upcoming quantum network architectures \cite{network}. No
complete homodyne reconstruction of the density matrix is needed
in such schemes. Very recently, a first important and promising step in this
direction has been realized with the experimental implementation of
direct photon detection for the measurement of the squeezing and purity
of a single-mode squeezed vacuum with a setup that required only a tunable beam splitter
and a single-photon detector \cite{wenger}.

As already anticipated, for $p=2$ we can provide analytical expressions
for the extremal entanglement in the space of global and
marginal purities
\bea
E_{{\N}max}(\mu_{1,2},\mu) = -\frac12\log \Bigg[-\frac{1}{\mu}
  + \left(\frac{\mu_1+\mu_2}{2\mu_1^2 \mu_2^2} \right)
  \nonumber \\
  \times  \left(\mu_1+\mu_2 -
  \sqrt{(\mu_1+\mu_2)^2-\frac{4 \mu_1^2 \mu_2^2}{\mu}} \right) \Bigg] \,
, \label{enmax}\\
E_{{\N}min}(\mu_{1,2},\mu) = - \frac12 \log
\Bigg[\frac{1}{\mu_1^2}+\frac{1}{\mu_2^2}-\frac{1}{2\mu^2} -
  \frac12  \nonumber\\
- \sqrt{\left( \frac{1}{\mu_1^2}+\frac{1}{\mu_2^2}-
\frac{1}{2\mu^2} - \frac12 \right)^2 - \frac{1}{\mu^2}} \; \Bigg]
\, . \label{enmin}
\eea
Consequently, both the average logarithmic
negativity $\delta \bar{E}_{\N}$, defined in \eq{average}, and the
relative error $\delta \bar{E}_{\N}$, given by
Eq.~(\ref{deltaen}), can be easily evaluated in terms of the
purities. The relative error is plotted in
Fig.~\ref{figerror2} for symmetric states as a function of the
ratio $S_{L_i} / S_L$. Notice, as already pointed out in the
general instance of arbitrary $p$, how the error decays exponentially.
In particular, it falls below $5\%$ in the range $S_L<S_{L_i}\ (\mu
> \mu_i)$, which excludes at most very weakly entangled states
(states with $E_{\N} \lesssim 1$ \cite{sqpar}). Let us remark that
the accuracy of estimating entanglement by the average logarithmic
negativity proves
even better in the nonsymmetric case $\mu_1 \neq \mu_2$,
essentially because the maximal allowed entanglement decreases
with the difference between the marginals, as shown in
Fig.~(\ref{gmemms})

The above analysis proves that the average logarithmic
negativity $\bar{E}_{\N}$ is a reliable estimate of the
logarithmic negativity $E_{\N}$, improving as the entanglement
increases. This allows for an accurate quantification of
continuous variable entanglement by knowledge of the global and
marginal purities. As we already mentioned, the latter quantities
may be in turn amenable to direct experimental determination by
exploiting recent single--photon detection proposals or the
upcoming technology of quantum networks.\par

\section{Summary and concluding remarks}\label{concl}

Summarizing, we have pointed out the existence of both maximally
and minimally entangled two--mode Gaussian states at fixed local
and global generalized $p-$entropies. The analytical properties of
such states have been studied in detail for any value of $p$.
Remarkably, for $p\le 2$, minimally entangled states are minimum
uncertainty states, saturating the Heisenberg principle, while maximally
entangled states are nonsymmetric two--mode squeezed thermal states.
Interestingly, for $p>2$ and in specific ranges of the values of
the entropic measures, the role of such states is reversed. In
particular, for such quantifications of the global and local
entropies, two--mode squeezed thermal states, often referred to as
continuous variable analog of maximally entangled states, turn
out to be \emph{minimally} entangled.

In the search for the extremally entangled states, we focused on
the hierarchy of mixedness induced by $p-$entropies on the set of
arbitrary multi--mode states, investigating several related
subjects, like the ordering of such different entropic measures
and the analytical comparison between the generic $S_p$
(and in particular the von Neumann
entropy $S_V$) and the linear entropy $S_L$.
Moreover, we have introduced the notion of ``average logarithmic
negativity'' for given global and marginal generalized $p$-entropies,
showing that it provides a reliable estimate of CV
entanglement in a wide range of physical parameters.

Our analysis also clarifies the reasons why the linear entropy is
a `privileged' measure of mixedness in continuous variable
systems. It is naturally normalized between $0$ and $1$, it offers
an accurate qualification and quantification of entanglement of
any mixed state while giving significative information about the
state itself and, crucially, is the only entropic measure which
could be directly measured in the near future by schemes involving
only single-photon detections or the technology of quantum networks,
without requiring a full homodyne reconstruction of the state.

More generally, the present analysis shows that some of
the canonical measures of entanglement and mixedness in the
discrete variable scenario, such as the entanglement of formation
and the von Neumann entropy, may not be the best choices for
the characterization of mixed states of continuous variable systems.
Discontinuous behaviors appear in the limit of infinite dimensions
for quantities like the entanglement of formation, even when
restricted to the special class of Gaussian states.
Entanglement measures such as the logarithmic negativity,
and informational measures such as the linear entropy and
the generalized entropies of higher order
provide quantifications which are, in many respects, more
satisfactory (both mathematically and physically) in the
context of CV systems.

\acknowledgments{We thank INFM, INFN, and MIUR under national
project PRIN-COFIN 2002 for financial support.}


\begin{thebibliography}{99}
\bibitem{foot1} We mention that, indeed, the continuous variable analogous of
maximally entangled states have been introduced in Ref.~\cite{keylwer}, by
considering singular states of the algebra of bounded operators.
\bibitem{sqpar} It is straightforwad to verify that, in the instance of two--mode squeezed thermal
(symmetric) states, such a condition corresponds to $\cosh(2r)\gtrsim \mu^{1/4}$. This
constraint can be easily satisfied with the present experimental technology: even for
the quite unfavorable case $\mu=0.5$ the squeezing parameter needed is just $r\simeq 0.3$.
\end{thebibliography}
\end{document}